\documentclass[aps,prl,showpacs,notitlepage,floatfix,twocolumn,superscriptaddress]{revtex4-2}
\usepackage{amssymb}
\usepackage{amsmath}
\usepackage{amsfonts}
\usepackage{bbm}
\usepackage{graphicx}
\usepackage{braket}
\usepackage{mathtools}
\usepackage{dsfont}
\usepackage{physics}
\usepackage{bbold}
\usepackage{enumerate}
\usepackage[titletoc]{appendix}

\usepackage[dvipsnames]{xcolor}

\newcommand\myshade{85}
\colorlet{myurlcolor}{MidnightBlue}

\usepackage[colorlinks,linkcolor=myurlcolor!\myshade!black,anchorcolor=myurlcolor!\myshade!black,citecolor=myurlcolor!\myshade!black,urlcolor=MidnightBlue]{hyperref}

\def\bea{\begin{equation}\begin{aligned}}
\def\eea{\end{aligned}\end{equation}}

\begin{document}
\setlength{\unitlength}{1mm}

\title{Noise Induced Universal Diffusive Transport in Fermionic Chains}
\author{Christopher M.~Langlett}
\email{clanglett85@tamu.edu}
\affiliation{Department of Physics \& Astronomy, Texas A\&M University, College Station, Texas 77843, USA}

\author{Shenglong Xu}
\affiliation{Department of Physics \& Astronomy, Texas A\&M University, College Station, Texas 77843, USA}

\begin{abstract}
We develop a microscopic transport theory in a randomly driven fermionic model with and without linear potential. The operator dynamics arise from the competition between noisy and static couplings, leading to diffusion regardless of ballistic transport or Stark localization in the clean limit. The universal diffusive behavior is attributed to a noise-induced bound state arising in the operator equations of motion at small momentum. By mapping the noise-averaged operator equation of motion to a one-dimensional non-hermitian hopping model, we analytically solve for the diffusion constant, which scales non-monotonically with noise strength, revealing regions of enhanced and suppressed diffusion from the interplay between onsite and bond dephasing noise, and a linear potential. For large onsite dephasing, the diffusion constant vanishes, indicating an emergent localization. On the other hand, the operator equation becomes the diffusion equation for strong bond dephasing and is unaffected by additional arbitrarily strong static terms that commute with the local charge, including density-density interactions.
The bound state enters a continuum of scattering states at finite noise and vanishes. However, the bound state reemerges at an exceptional-like point in the spectrum after the bound-to-scattering state transition. We then characterize the fate of Stark localization in the presence of noise.
\end{abstract}
\maketitle

\begin{figure}[t]
    \includegraphics[width=\columnwidth]{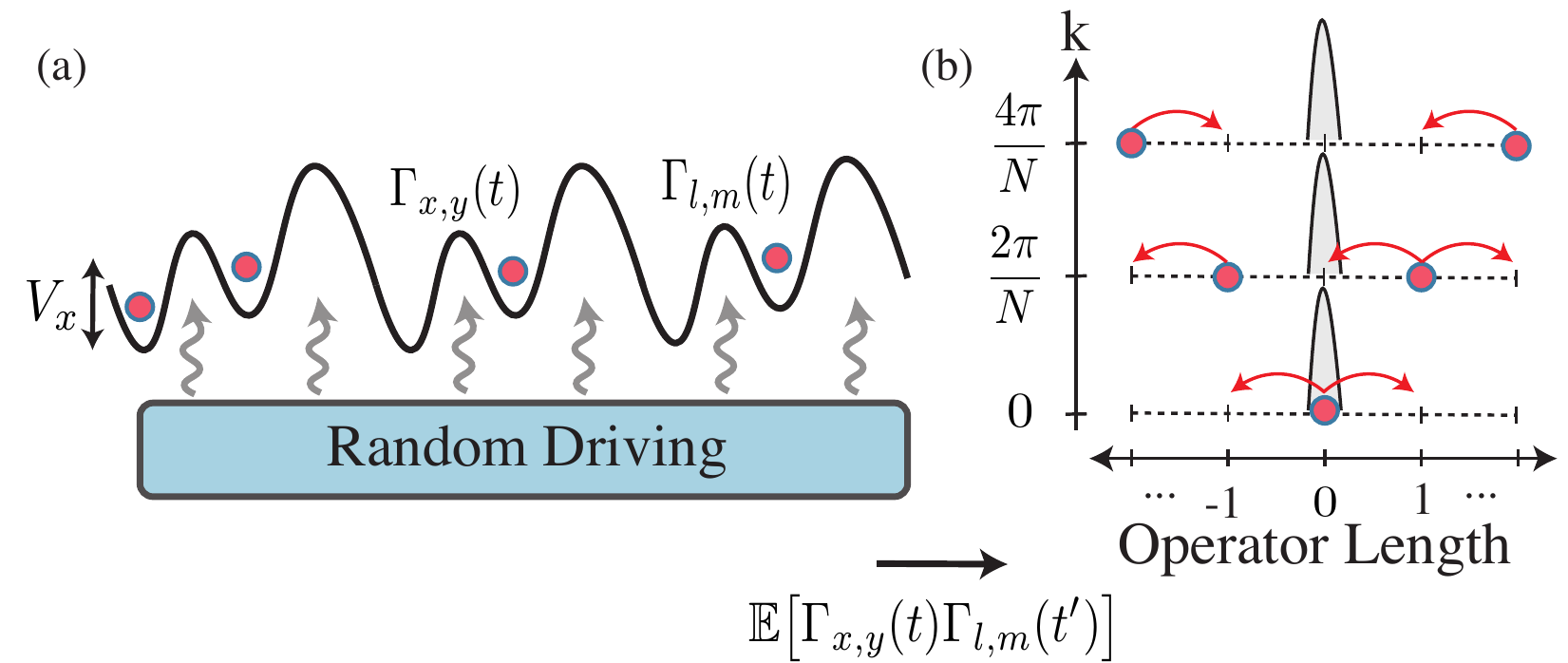}
    \caption{\textit{Noise Induced Non-Hermitian Hopping Model.}
    (a) Randomly driven non-interacting fermions in a spatially dependent potential, $V_{x}$. Classical noise $\Gamma_{x,y}(t)$ models the random drive by coupling locally to the hopping or density.
    (b) Noise-averaged operator equations of motion map onto a set of one-dimensional non-hermitian hopping models with a repulsive delta function. The $x$-axis is the operator length, and $k$ is the center-of-mass momentum.
    }
    \label{fig:Fig1}
\end{figure}

An outstanding challenge of many-body physics is a complete explanation of how phenomenological laws governing irreversible macroscopic transport behavior emerge from reversible microscopic dynamics, a process encapsulated by the eigenstate thermalization hypothesis~\cite{d2016quantum,ChaosSrednicki1994,QuantumDeutsch1991}. This challenge only magnifies in interacting quantum many-body systems in both equilibrium and non-equilibrium processes~\cite{ColloquiumPolkovnikov2011,eisert2015quantum}.
Along these lines, one-dimensional systems~\cite{giamarchi2003quantum,FermiGuan2013} are attractive because quantum fluctuations have a pronounced effect, leading to a wide array of quantum phenomena ranging from ballistic transport to localization.
In particular, the observation of superdiffusive transport~\cite{KardarLjubotina2019, ljubotina2017spin,spohn20201,scheie2021detection,wei2022quantum,Anomalousde2019,Superuniversality} beyond the expected ballistic behavior in integrable systems. However, a complete characterization of quantum transport in solvable models remains challenging despite having access to the eigenenergies and excitations~\cite{bethe1931theorie}.

Randomly driven models, in which couplings are random variables uncorrelated in time, help understand the spreading of a local operator under Heisenberg evolution, known as the operator dynamics.
Systems with added stochasticity ought to lose their microscopic properties, such as conservation laws, permitting the emergence of universal behavior.
These systems have recently been revitalized with discrete time evolution involving dual unitary circuits~\cite{ExactPiroli2020,ExactBertini2019} and replica disorder averaged random unitary circuits~\cite{fisher2023random,OperatorNahum2018,harrow2009random}.
On the other hand, stochastic dynamics of continuous time models in random Hamiltonians~\cite{shenker2015stringy,sunderhauf2019quantum,lashkari2013towards,Localityxu2019,saad2018semiclassical,OperatorZhou2019}, noisy spin chains~\cite{EntanglementKnap2018, NoisyRowlands2018, NoiseGopalakrishnan2017, DrivenSingh2021, swann2023spacetime, fava2023nonlinear}, and (a)symmetric simple exclusion processes~\cite{From2020Jin,Open2019Bernard,DynamicsDenis2022,eisler2011crossover} have provided deep insights.
Random unitary dynamics have also attracted experimental interest in cold atoms~\cite{bhatt2022stochastic,shimasaki2022anomalous,ObservationSajjad2022}, trapped ions~\cite{bermudez2010localization,noel2022measurement,MaierEnvironment2019}, and paraxial optics~\cite{levi2012hyper}.

Despite tremendous progress, a complete characterization of the ingredients necessary for unorthodox transport to arise in interacting many-body systems remains open.
One approach is introducing a static term as a perturbation~\cite{NoiseGopalakrishnan2017, StabilityNardis2021} to access more generic information about late-time transport. A recent study~\cite{AbsenceClaeys2022} of a spin-1/2 chain with exchange couplings that fluctuate in space-time around a non-zero mean revealed, through perturbation theory, late-time spin diffusion, albeit with a superdiffusive enhancement suggesting normal diffusion~\cite{glorioso2021hydrodynamics}.

In this work, we extend these results to non-perturbative static terms. We develop a microscopic transport theory in a fermionic chain without and in the presence of a linear potential. In both cases, the operator dynamics arise from the competition between randomly driven and arbitrarily strong static couplings. We analytically solve for the diffusion constant by exactly mapping the noise-averaged operator equation of motion to a one-dimensional non-hermitian hopping model—the diffusion constant scales non-monotonically with noise strength, revealing enhanced and suppressed diffusion regions. 

We uncover for all noise models that a diffusive mode governs the late-time hydrodynamics at small $k$, attributed to an emergent bound state in the operator equations of motion. As $k$ increases, the bound state enters a scattering state continuum and vanishes. From the non-hermitian structure of the operator equations, the bound state reemerges at an exceptional-like point where a pair of complex energies form. However, for strong bond dephasing noise, the operator equation becomes the diffusion equation and is \textit{unaffected} by additional arbitrarily strong static terms that commute with the local charge, including density-density interactions. Moreover, we then characterize the fate of Stark localization in the presence of noise. Ultimately, noise destabilizes the Stark ladder, allowing transport to occur albeit non-monotonically.

\textit{Model.}---
We explore the dynamics of one-dimensional non-interacting fermions with time-dependent noise~\cite{Classical2009Amir,Exact2016Medvedyeva}, through the Hamiltonian,
\begin{equation}
\label{eq:freeferm}
    H_{t} = \sum_{x, y}\big[J_{x,y}+\Gamma_{x,y}(t) \big]c_{x}^{\dagger}c_{y},
\end{equation}
where $c^{\dagger}_{x}~(c_{x})$ create~(annihilate) an electron at site index $x$.
The off diagonal elements of $J_{x,y}$ and $\Gamma_{x,y}(t)$ represent either static or driven hopping, while the diagonal elements represent a static or driven potential.
The amplitudes $\{\Gamma_{x,y}\}$ are drawn independently for each pair of sites~($x,y$) from a Gaussian distribution with zero mean and variance,
\begin{align}
\label{eq:coupling}
\mathbb{E}[\Gamma_{x,y}(t)\Gamma_{l,m}(t^\prime)] & = \Gamma_{xy}\delta_{x,l}\delta_{y,m}\delta(t-t^\prime).
\end{align}
Where $\mathbb{E}[\cdot]$ denotes the average over disorder, $\Gamma_{x,y}$ sets the energy scale of the noise, and $\delta(t-t^\prime)$ implies the couplings are correlated at a single instance in time.

We study analytically and numerically time-dependent correlation functions to reveal the long-distance late-time hydrodynamic transport in the presence of noise.
In the Heisenberg picture, the infinitesimal operator evolves stochastically, $\mathcal{O}_{t+dt}=e^{iH_{t}dt}\mathcal{O}_{t}e^{-iH_{t}dt}$.
The evolution equation for a generic noise-averaged operator follows from expanding the flow of $\mathcal{O}_{t}$ up to second-order in $dt$ and averaging the noise~\cite{hudson1984quantum,bauer2014open,gardiner2004quantum},
\begin{equation}
\label{eq:Motion}
d\bar{\mathcal{O}_{t}} = \sum_{x,y} \bigg[ iJ_{x,y}[c^{\dagger}_{x}c_{y}, \bar{\mathcal{O}_{t}}] + \Gamma_{x,y} \mathcal{L}_{x,y}[\bar{\mathcal{O}_{t}}] \bigg] dt.
\end{equation}
Here the average dynamics are governed by an effective Lindblad description~\cite{vznidarivc2010exact,Universal2023Alexios,Relaxation2015marko,Open2019Bernard} where $\mathcal{L}_{x,y}[\boldsymbol{*}]=L^{\dagger}_{x,y}\boldsymbol{*} L_{x,y}-\frac{1}{2}\{L^{\dagger}_{x,y}L_{x,y},\boldsymbol{*} \}$ with $L_{x,y}=c^{\dagger}_{x}c_{y}+h.c$, and $\{,\}$ standing for the anti-commutator~\footnote{We note that the form of the above equation follows a convention different than standard condensed matter transport papers where the equations of motion are set up to resemble the Schrodinger equation i.e., $i d\mathcal{O}=[H, \mathcal{O}]dt$. As such the eigenvalues of Eq.~\eqref{eq:Motion} are purely imaginary without noise and the real part represents decoherence.}.
Competition between coherent and incoherent dynamics drive the time evolved noise-averaged operator in the late-time limit to the steady state $\lim_{t\to\infty}\bar{\mathcal{O}_{t}}=\sum_{x}n_{x}$ from charge conservation.

\textit{Characterizing Transport.}---
Universal behavior of the random unitary dynamics is ascertained through the infinite-temperature fermion density-density correlation function,
\begin{equation}
    \label{eq:correlator}
    C_{x,y}(t) = \frac{1}{2^{N}}\tr [\left(n_{x}(t)-\frac{1}{2}\right)\left(n_{y}-\frac{1}{2}\right)],
\end{equation}
where $n_{x}(t)$ denotes the time-evolved density operator at site index $x$ in the Heisenberg picture.
The density-density correlation function Eq.~\eqref{eq:correlator} decays with an algebraic tail at late times,
\begin{equation}
    \label{eq:latetimecorrelator}
    \lim_{t\to\infty} \lim_{N\to\infty} C_{N/2,N/2}(t) \sim t^{-1/z}.
\end{equation}
The dynamical exponent $z$ classifies the universal hydrodynamic transport behavior, for example, $z=1$ for ballistic, $1<z<2$ for superdiffusive, $z=2$ for diffusive, $z>2$ is subdiffusive, and $z=\infty$ for localized.

\begin{figure}[t]
    \includegraphics[width=\columnwidth]{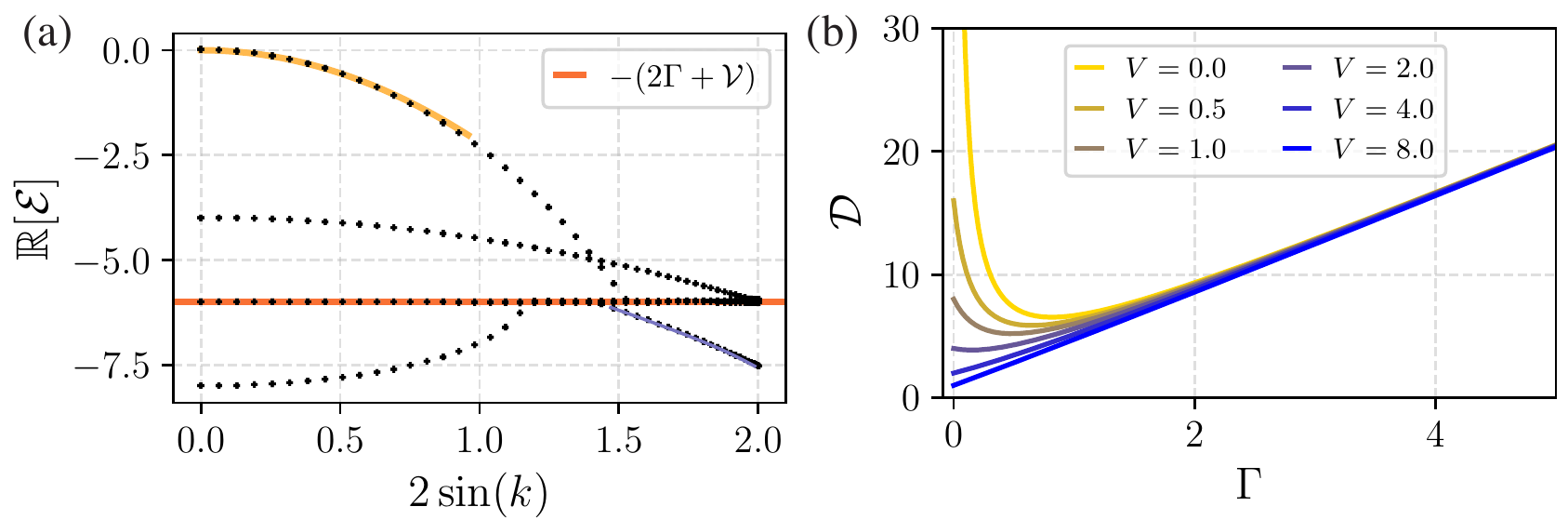}
    \caption{\textit{Bond and Onsite Dephasing Noise.}
      (a) Real part of eigenvalue spectrum with both onsite and bond dephasing noise. The yellow curve is the diffusive mode corresponding to Eq.~\eqref{eq:Case1BSE}. The red line indicates the continuum of scattering states, and the blue curve is a degenerate set of complex energies.
      (b) Diffusion constant from Eq.~\eqref{eq:Case1BSE}. When $\Gamma=0$ the diffusion constant decreases from a ballistic~($\mathcal{V}\rightarrow 0$) to an emergent localization regime when $\mathcal{V}\rightarrow \infty$. As $\Gamma$ reaches the minimum $\sqrt 6 J -\mathcal{V}$ then \textit{increases} monotonically into a noise-assisted transport regime.
      Parameters: (a) $N=400$, $\gamma = 0$, $\Gamma/J=2$, $\mathcal{V}/J=2$.
    }
    \label{fig:Fig2}
\end{figure}

\textit{Operator Dynamics.}---
The Heisenberg operator $n_{x}(t)$ remains a two-body operator under evolution due to the absence of interactions, permitting the expansion,
\begin{equation}
\label{eq:OpExp}
n_{x}(t) = \sum_{m,n=1}^{N} A_{m,n}(t)c^{\dagger}_{m}c_{n}.
\end{equation}
With the initial condition, $A_{m,n}(0)=\delta_{m,x}\delta_{n,x}$.
We transform into the coordinates $\ell=n-m$~\footnote{To perform numerical simulations it is necessary to set an operator cutoff length we denote $\ell_{\text{max}}$ which we always set to $N/2$ where $N$ is the system size.} and $\mathcal{R}=n+m$ representing the operator length and center-of-mass.
Because the noise-averaged operator equation is translation invariant in $\mathcal{R}$ in our models, a Fourier transformation maps Eq.~\eqref{eq:Motion} to equations for $A_{\ell,k}$ describing a one-dimensional hopping model on a fictitious lattice of operator length $\ell$ with the center of mass momentum $k$~[see Fig.~\ref{fig:Fig1}]. 
The correlation function, in terms of the coefficients is given by, 
$
\frac{1}{8\pi }\int dk A_{0, k}(t) e^{i k (x-y)}
$
, where $A_{\ell,k}(t)$ is the time-evolved wavefunction of the effective hopping model and $A_{\ell,k}(0)=\delta_{\ell,0}$. 
At finite noise, the effective model is non-Hermitian, where the non-positive real parts of the eigenvalues drive the system to the steady state in the late-time limit, corresponding to the eigenvalue with the maximal real part, namely, the eigenstate decays slowest during time evolution. 

\textit{Bond and Onsite Dephasing Noise.}---
We now focus our model in Eq.~\eqref{eq:freeferm} on nearest-neighbor hopping with dephasing noise on both bonds and sites.
Specifically, we define the parameters,
\begin{equation}
    J_{x,x+1}=J,\quad \Gamma_{x,x} = \mathcal{V},\quad \Gamma_{x,x+1}=\Gamma.
\end{equation}
Here $J$ is the nearest-neighbor coherent hopping, $\mathcal{V}$ and $\Gamma$ are the onsite and bond dephasing strength, respectively.  
The eigenvalue equations of Eq.~\eqref{eq:Motion} take the form 
\begin{align}
\label{eq:Case1}
\mathcal{E}_{q} A_{0}  &= t_{k}\big[A_{1} - A_{-1}\big]-4\Gamma \sin^{2}(k)A_{0}\nonumber\\
\mathcal{E}_{q} A_{\pm 1}  &= \pm t_{k}\big[A_{\pm 2} - A_{0}\big]+\Gamma A_{\mp 1}-\big[\mathcal{V}+2\Gamma\big] A_{\pm 1}\nonumber \\
\mathcal{E}_{q} A_{\ell}  &= t_{k}\big[A_{\ell+1} - A_{\ell-1}\big]-\big[\mathcal{V}+2\Gamma\big] A_{\ell}.
\end{align}
We dropped the index $k$ in $A_{\ell,k}$ for simplicity, and $q$ labels different levels of the eigenvalue equation.
The first two equations are the boundary conditions near the origin of the fictitious operator length lattice, and the third describes the bulk for $|\ell|>1$ with the effective hopping, $t_{k}=2J\sin(k)$.
There are two well known limits of Eq.~\eqref{eq:Case1}; no noise, $\Gamma = \mathcal{V}=0$, and pure dephasing, $J=0$.
In the former case, the model is purely coherent, leading to the correlation function,
\begin{equation}
\label{eq:BesselFunc}
    C_{x,y}(t) = \frac{1}{4}\mathcal{J}_{x-y}^{2}(2Jt).
\end{equation}
Here $\mathcal{J}_{x-y}(2Jt)$ is the Bessel function of the first kind of order $x-y$.
The asymptotic behavior of the correlation function, $\lim_{t\to\infty}C_{N/2,N/2}(t)=1/ \pi t$, indicates ballistic transport with an exponent $z=1.0$.
In the latter case~($J=0$ or equivalently $t_k=0$), the operator length $\ell=0$ decouples from all other operator lengths, mapping to the diffusion equation, with the solution,
\begin{equation}
\label{eq:ModBesselFunc}
    C_{x,y}(t) = \frac{1}{4}e^{-2 \Gamma t}\mathcal{I}_{x-y}(2\Gamma t).
\end{equation}
Here $\mathcal{I}_{x-y}(2\Gamma t)$ is the modified Bessel function of the first kind of order $x-y$.
The asymptotic scaling of Eq.~\eqref{eq:ModBesselFunc} is, $\lim_{t\to\infty} C_{N/2,N/2}(t)=1/ 2\sqrt{t\pi}$ corresponding to the exponent $z=2$.
Including a static potential that couples to the density do not affect the diffusive mode because it commutes with the local charge $n_{x}$ and bond dephasing leaves $n_{x}$ unchanged.
Generically, including any static term that commutes with the local charge, even the density-density interaction, $n_x n_y$, will not affect the diffusive hydrodynamic mode.

Now we solve Eq.~\eqref{eq:Case1} for general $J$, $\mathcal{V}$ and $\Gamma$. It is similar to the standard Schr\"{o}dinger equation with a $\delta$ potential; both scattering and bound states exist in the spectrum, whereby the bulk equation fixes the real part of the scattering states energy to be $-\big[\mathcal{V}+2\Gamma\big]$~[see red line in Fig.~\ref{fig:Fig2}(a)].
Translation invariance of Eq.~\eqref{eq:Case1} permits the ansatz,
\begin{equation}
\label{eq:Case1Ansatz}
    A_{\ell} = \begin{cases}
    A_{-1}e^{q(1+\ell)} & \text{if}\ \ell \leq -1\\
    -A_{1}e^{q(1-\ell)+i\pi\ell}& \text{if}\ \ell \geq 1.
    \end{cases}
\end{equation}
Inserting the above solution into the bulk equation, gives the energy, $\mathcal{E}_{q}=4\sin(k)\sinh(q)-\mathcal{V}-2\Gamma$.
The boundary conditions for $|\ell|\leq 1$ constraint the values of $q$ through~[see SM~\cite{SupMat}],
\begin{equation}
\label{eq:Casse1Cubic}
    \big[\mathcal{E}_{q}+4\Gamma \sin^{2}(k) \big] \big[ t_{k}e^{q} + \Gamma\big] = -2t^{2}_{k}.
\end{equation}
The above equation is an exactly solvable cubic equation, which at small $k$ admits two physical solutions, one that begins at $\mathcal{E}_{q} = 0$~[see yellow curve in Fig.~\ref{fig:Fig2}(a)] and the other at $\mathcal{E}_{q}=-[3\Gamma + \mathcal{V}]$~[lowest branch in Fig.~\ref{fig:Fig2}(a)].
The branch in Fig.~\ref{fig:Fig2}(a) beginning at $\mathcal{E}_{q}=-[\Gamma + \mathcal{V}]$ is determined by solving Eq.~\eqref{eq:Case1} assuming $A_{0}=0$.
Moreover, the gapless bound state energy is given by,
\begin{equation}
    \label{eq:Case1BSE}
    \mathcal{E}_{q} = -4\bigg[\Gamma + \frac{2J^{2}}{\mathcal{V}+3\Gamma}\bigg]k^{2}.
\end{equation}
A diffusive mode always exists at small momentum regardless of whether the sites or the hopping have finite dephasing~[see the yellow curve in Fig.~\ref{fig:Fig2}(a)].
When both $\mathcal{V}, \Gamma \rightarrow 0$, the diffusion constant diverges, which is reminiscent of ballistic transport in the coherent limit.Previously obtained was the result with either only onsite or bond dephasing noise~\cite{eisler2011crossover,esposito2005exactly}.
In general, the diffusion constant decreases monotonically with increasing onsite dephasing $\mathcal{V}$ because an energy barrier from site-to-site impedes coherent hopping.
In particular, in the absence of bond dephasing, the diffusion constant is zero in the large $\mathcal{V}$ limit, indicating an emergent localization. 
As illustrated in Fig.~\ref{fig:Fig2}(b), the diffusion constant displays non-monotonic behavior as a function of bond dephasing $\Gamma$.
Specifically, as $\Gamma$ increases, the diffusion constant reaches a minimum at $\Gamma=(\sqrt 6 J-\mathcal{V})/3$~(assuming $\mathcal{V}<\sqrt 6 J$), and then \textit{increases} monotonically, entering a regime of noise-assisted transport~\cite{rebentrost2009environment, MaierEnvironment2019,EffectsZerah2020}.  

As momentum increases, two interesting characteristics become apparent. First, the diffusive mode undergoes a bound-to-scattering state phase transition upon entering a scattering state continuum at $\mathcal{E}_{q}=-2\Gamma - \mathcal{V}$. Then, from the non-hermitian characteristic of Eq.~\eqref{eq:Case1}, there is an exceptional-like point~\cite{Spectral2019can, Decoherence2022Chen} where the two physical solutions of Eq.~\eqref{eq:Casse1Cubic} collide and coalesce, becoming a complex conjugate pair of energies visualized by the doubly degenerate points in Fig.~\ref{fig:Fig2}(a) indicated with a blue curve. 

\begin{figure}[t]
    \includegraphics[width=\columnwidth]{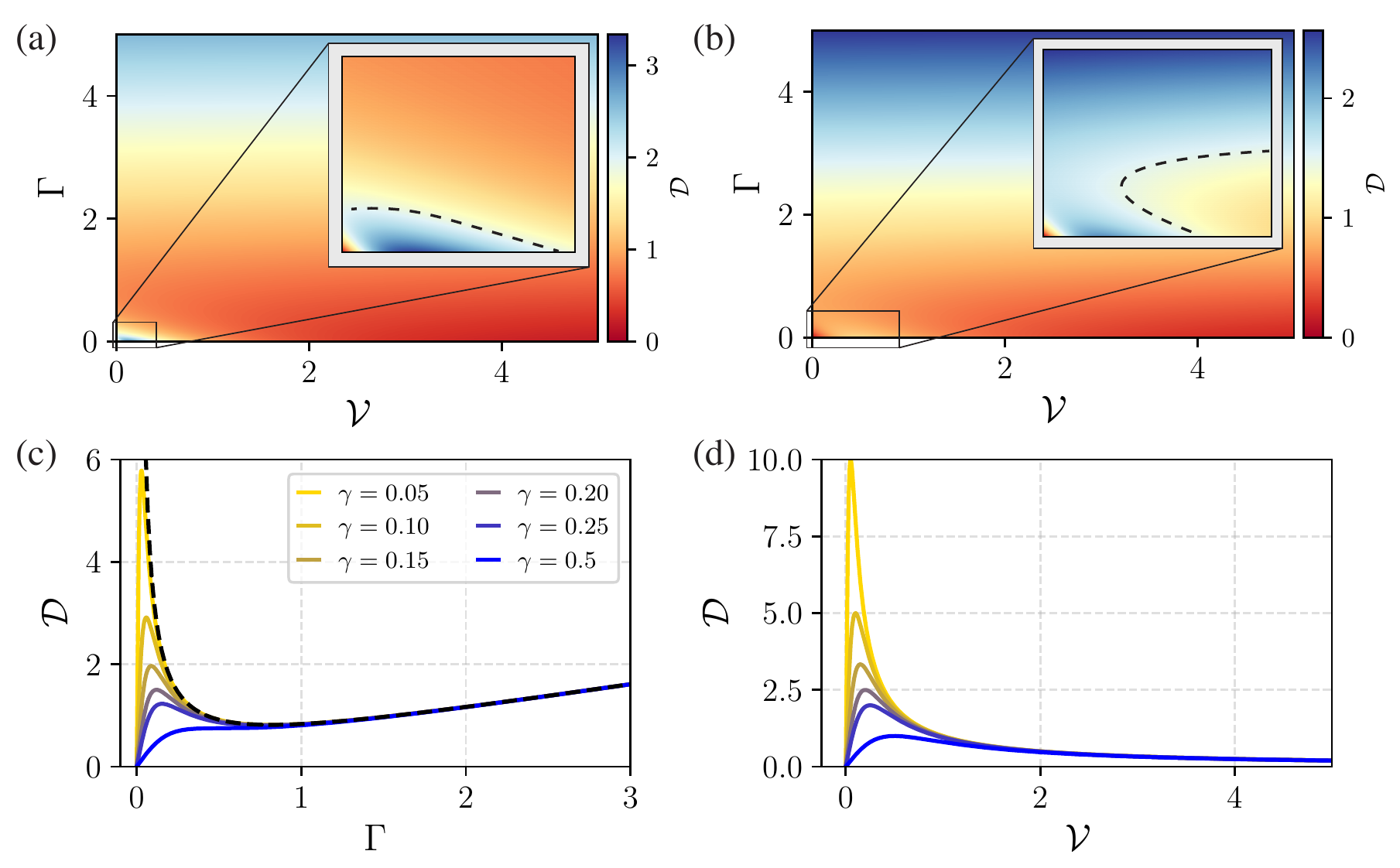}
    \caption{\textit{Diffusion Constant Phase Diagram.}
      (a) Diffusion constant from Eq.~\eqref{eq:FullBSE} with the linear potential strength $\gamma = 0.15$. Inset: Illustration of the non-monotonicity along both axes.  
      (b) Same as in (a) with $\gamma = 0.50$ where the non-monotonic behavior arises only along $\Gamma = 0$. Inset: Illustration of non-monotonicy along the onsite dephasing axis only.
      (c) Diffusion constant with $\mathcal{V} = 0$. Provided $\gamma < 0.5$ there is an initial noise assisted regime to a maximum value, where then bond dephasing introduces an energy barrier, suppressing diffusion. Once $\Gamma > \gamma$ diffusion enhances as if the linear potential was absent~[see black curve for $\gamma=0$ or Fig.~\ref{fig:Fig2}(b)]. As $\gamma \rightarrow 0.5$ the non-monotonic behavior is lost, and diffusion immediately enters a noise-assisted transport regime. 
      (d) While when $\Gamma = 0$ noise compensates for the energy barrier from the linear potential, enhancing transport to a maximum. As $\mathcal{V}$ increases further, the onsite dephasing dominates the linear potential, introducing an energy barrier and decreasing the diffusion constant. 
      Parameters: The dotted black curves in (a) and (b) indicate a maximum or minimum.
    }
    \label{fig:Fig3b}
\end{figure}

\textit{Linear Potential with Bond and Onsite Dephasing.}---
In the clean limit of the previous examples, the system exhibited ballistic transport~[see Eq.~\eqref{eq:BesselFunc}].
However, no matter how weak or the location, finite noise causes diffusive transport. We now turn our attention to the opposite limit, where in the clean limit, the system is localized, and the diffusion constant vanishes.
We will study Wannier-Stark localization in the presence of noise~\cite{Effect2019Bhakuni, Effect2020Bandyopadhyay,Dephasing2007Yoshida,BathWu2019}.
Specifically, consider the linear potential $J_{x,x}=-\gamma x$ where $\gamma$ is the slope with the noise coupled to the hopping and density.
We now study the competition between these two noise models through the equation,
\begin{equation}
    \label{eq:FullNoise}
    \mathcal{E}_{q} A_{\ell, k}=t_{k}\big[A_{\ell+1, k}-A_{\ell-1, k}\big] +\big[i\gamma \ell-2\Gamma-\mathcal{V}\big]A_{\ell, k}.
\end{equation}
The bulk operator equation is no longer translation invariant in $\ell$, which permitted the plane wave ansatz Eq.~\eqref{eq:Case1Ansatz}.
Solving the recursion relation, $A_{\ell}$ instead takes the form,
\begin{equation}
    \label{eq:StarkBoundStateAnsatz}
    A_{\ell, k} = \begin{cases}
    A\mathcal{I}_{\nu_{-}}(-2it_{k}/\gamma) & \text{if}\ \ell < -1\\
    B\mathcal{I}_{\nu_{+}}(-2it_{k}/\gamma) & \text{if}\ \ell >1.
    \end{cases}
\end{equation}
where $\nu_{\pm}=i(\mathcal{E}_{q} +2\Gamma+\mathcal{V})/\gamma\pm \ell$.
For $\mathcal{V}=\Gamma=0$ the operator equations are anti-hermitian leading to an equally spaced tower of purely imaginary eigenvalues, $\mathcal{E}_{q}=i\gamma q$ for $q \in \{ -\ell_{\text{max}}, \ell_{\text{max}}\}$ independent of momentum $k$.
The corresponding unnormalized eigenstates are $A_{\ell,k}=\mathcal{I}_{\ell-q}(-4iJ\sin(k)/\gamma)$ which are Wannier-Stark localized~\cite{DynamicsWannier1962,ExistenceEmin1987, StarkMendez1988,schmidt2018signatures}.
Finite noise renders the operator equations non-hermitian, causing an eigenvalue to become purely real, which is the long wavelength mode.
In the SM~\cite{SupMat}, we determine the scaling of the hydrodynamic mode,
\begin{equation}
\label{eq:FullBSE}
    \mathcal{E}_{q} = -8 \bigg[\frac{\Gamma}{2} +\frac{J^{2}(\mathcal{V}+\Gamma)}{\gamma^{2} +(\mathcal{V}+\Gamma)(\mathcal{V}+3\Gamma)}\bigg]k^{2},
\end{equation}
which is diffusive for finite noise, similar to Anderson localized models with global noise~\cite{ClassicalAmir2009,NoiseGopalakrishnan2017,evensky1990localization}, but different from local noise models~\cite{RemnantsLorenzo2018,lm2022logarithmic}. In the limit $\gamma=0$, we recover the bound state energy Eq.~\eqref{eq:Case1BSE}, while in the limit either $\mathcal{V}$ or $\Gamma$ is large, the bound state energy is finite, specifically, $4\Gamma$, indicating Stark localization instability to noise.

In Fig.~\ref{fig:Fig2}(a) and (b), we plot the heatmap of the diffusion constant with $\gamma<0.5$ and $\gamma=0.5$. In both cases, the model is Stark localized when $\mathcal{V}=\Gamma=0$. When $\gamma<0.5$, initially, there is a regime where increasing $\Gamma$ or $\mathcal{V}$ leads to noise-assisted transport to a maximum value~[see Fig.~\ref{fig:Fig2}(c) or (d)]. Increasing noise further in either direction introduces an energy barrier that overcomes the linear potential, suppressing diffusion; however, when $\Gamma>\gamma$, diffusion enhances once more as if the linear potential was nonexistent~[see the black curve for $\gamma=0$ in Fig.~\ref{fig:Fig3b}(c) or Fig.~\ref{fig:Fig2}(b)]. As $\gamma \rightarrow 0.5$, the non-monotonic behavior decreases and is lost when $\gamma >0.5$, whereby diffusion immediately enters a noise-assisted transport regime. On the other hand, the onsite dephasing dominates the linear potential as $\mathcal{V}$ increases~[see Fig.~\ref{fig:Fig3b}(d)], introducing an energy barrier and decreasing the diffusion constant.

\begin{figure}[t]
    \includegraphics[width=\columnwidth]{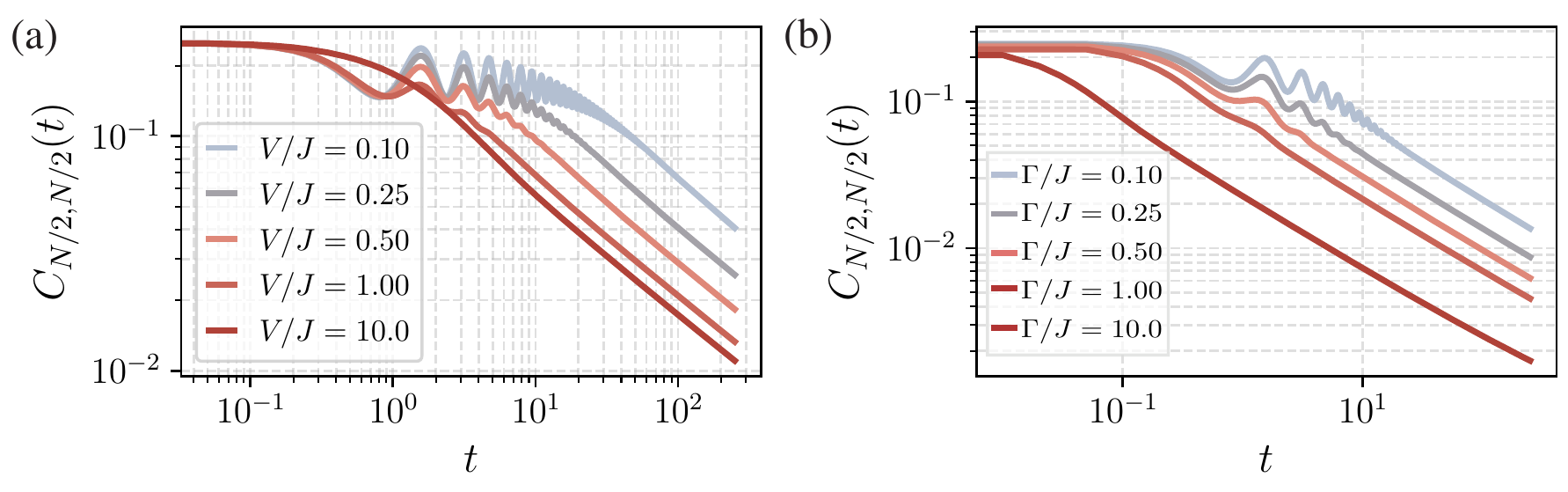}
    \caption{\textit{Noisy Linear Potential Operator Dynamics.}
      (a) The auto-correlation function $C_{N/2,N/2}(t)$ with onsite dephasing. The oscillating behavior is a signature of the underlying Stark localization, which pushes the onset of diffusion to late times.
      (b) Same as (a) but bond dephasing noise.
      Parameters: (a) and (b): $N=400$, $dt=0.05$, $\gamma=4$, $J=1$.
    }
    \label{fig:Fig3}
\end{figure}

We first study the operator dynamics of Eq.~\eqref{eq:FullNoise} with only onsite dephasing present, i.e., $\Gamma=0$. When $\mathcal{V} \ll \gamma $ the diffusion constant is small, and Bloch oscillations push diffusion to later times~[see Fig.~\ref{fig:Fig3}(a)], rather than when $\mathcal{V}$ is the dominant energy scale. In contrast, diffusion almost immediately occurs when the noise is on the bonds~[see Fig.~\ref{fig:Fig3}(b)], i.e., $\mathcal{V}=0$; a consequence of the diffusion constant always being finite regardless of the linear potential strength.

\textit{Conclusion.}---
Through a combination of analytics and large-scale numerics, this work developed a transport model where the operator dynamics arise from the competition between randomly driven and static couplings. We exactly solve for the diffusion constant by determining the emergent bound state of an effective one-dimensional non-hermitian hopping model. In contrast to standard hydrodynamic theories~\cite{kac1956foundations,chapman1990mathematical}, the diffusion constant scales non-monotonically with noise strength. For pure dephasing, the noise-averaged equation satisfies the diffusion equation, which is robust to arbitrarily strong static terms that commute with the local charge, including interactions. As momentum increases, the bound state enters a continuum of scattering states and vanishes. Surprisingly, beyond the bound-to-scattering state phase transition, the bound state reemerges at an exceptional-like point. We further find Stark localization is unstable to onsite and bond dephasing noise, but illustrates a rich phase diagram where diffusion enters regimes of enhancement and suppression. Future work could be understanding transport when the model has long-range hopping or correlating the noise~\cite{AndersonMarcantoni2022}.

\textit{Acknowledgement.}--- We thank Lakshya Agarwal, Joaquin F. Rodriguez-Nieva, and Artem Abanov for useful discussions. We also thank Mark Mitchison for pointing out related results from previous works.
The numerical simulations in this work were conducted with the advanced computing resources provided by Texas A\&M High Performance Research Computing.

\bibliography{References}
\end{document}


\title{Supplemental Material for Noise Induced Universal Diffusive Transport in Fermionic Chains}
\author{Christopher M. Langlett}
\affiliation{Department of Physics \& Astronomy, Texas A\&M University, College Station,Texas 77843, USA}
\author{Shenglong Xu}
\affiliation{Department of Physics \& Astronomy, Texas A\&M University, College Station,Texas 77843, USA}

\begin{abstract}
This supplemental material's purpose is to clarify the conclusions drawn in the main text. We first give an overview of calculating the two-point correlation function utilizing free fermion techniques. Next, we derive the operator equation of motion that leads to the effective Liouvillian. In the third section, we explicitly derive the differential equation for the operator amplitudes. Solving the diffusion constant for different noisy and static combinations constitutes the remaining sections. The final section outlines the numerical methods for simulating the differential equations.
\end{abstract}
\maketitle
\tableofcontents

\section{Fermion Operator Identities.}
\label{appsec:ident}

In this appendix, we provide helpful fermion operator identities used to derive the equations of motion.
We denote the spinless fermionic creation and annihilation operators by $(c^{\dagger}_{\alpha}, c_{\alpha})$, where $\alpha$ indicates the site index.
These operators obey the algebra,
\begin{align}
\begin{split}
    \label{eq:FermAlg}
    \{c_{\alpha},c_{\beta} \}=\{c^{\dagger}_{\alpha},c^{\dagger}_{\beta} \}=0\\
    \{c^{\dagger}_{\alpha},c_{\beta} \}=\delta_{\alpha,\beta}.
    \end{split}
\end{align}
Further, defining the density operator $n_{\alpha}=c^{\dagger}_{\alpha}c_{\alpha}$, we obtain,
\begin{align}
\label{eq:FermComms}
    [n_{\alpha}, c_{\beta}] & = -\delta_{\alpha, \beta}c_{\beta},\quad [n_{\alpha}, c^{\dagger}_{\beta}] = \delta_{\alpha, \beta}c^{\dagger}_{\beta}\nonumber \\
    \{n_{\alpha}, c^{\dagger}_{\beta}\} & = \delta_{\alpha, \beta}c_{\alpha}^{\dagger} + 2c_{\beta}^{\dagger}n_{\alpha},\quad \{n_{\alpha}, c_{\beta}\} = \delta_{\alpha, \beta}c_{\alpha} + 2n_{\alpha}c_{\beta}
\end{align}
These relations were utilized in deriving the differential equations for the coefficients in the operator expansion.

\section{Free-Fermion Techniques.}
\label{appsec:FreeFermMeth}
We now calculate the density-density correlation function using standard methods. In the clean limit, $\Gamma_{x,y}=0$, with $J_{x,x}=0$, the exchange couplings are uniform, and the Hamiltonian describes a one-dimensional free-fermion.
At this point, the Hamiltonian regains spatial and time translation invariance, thereby in Fourier space, the fermionic operators become,
\begin{align}
\begin{split}
    c_{x} & = \frac{1}{\sqrt{N}}\sum_{k}e^{ik \cdot x}c_{k}\\
    c^{\dagger}_{x} & = \frac{1}{\sqrt{N}}\sum_{k}e^{-ik \cdot x}c^{\dagger}_{k}\\
\end{split}
\end{align}
here $k$ is the Brillouin zone momentum, and the Hamiltonian,
\begin{equation}
    \label{eq:MomFreeFerm}
    H = \sum_{k}\mathcal{E}_{k}c_{k}^{\dagger}c_{k}
\end{equation}
with the band dispersion, $\mathcal{E}_{k}=2\cos(k)$~(lattice spacing $a=1$).
In the momentum basis, time-evolution of the operators $c^{\dagger}_{k}~(c_{k})$ is simple, taking the form,
\begin{align}
    \begin{split}
    c_{k}(t) & = e^{-i\mathcal{E}_{k}t}c_{k}\\
    c^{\dagger}_{k}(t) & = e^{i\mathcal{E}_{k}t}c^{\dagger}_{k}.\\
    \end{split}
\end{align}
In real space, we have that $c_{x}(t)=\sum_{y}u(x-y, t)c_{y}$.
The function, $u(x-y, t)$, is the Fourier transform of the phases $e^{-i\mathcal{E}_{k}t}$, and controls how the correlation function grows in time.
Specifically, the envelope function takes the form,
\begin{equation}
    \label{eq:RealSpace}
    c_{x}(t) = \sum_{y}\underbrace{\bigg[\frac{1}{N}\sum_{k}e^{i(x-y)k} e^{-i\mathcal{E}_{k}t}\bigg]}_\text{$u(x-y, t)$}c_{y}.
\end{equation}
We now study the universal properties of this function.
In the thermodynamic limit, the mode functions take the form,
\begin{align}
\begin{split}
    \label{eq:envelope}
    u(x-y, t) & = \int_{-\pi}^{\pi}\frac{dk}{2\pi}e^{i(x-y)k}e^{-2i\cos(k)t}\\
    & = i^{(x-y)}\mathcal{J}_{x-y}(2t).
\end{split}
\end{align}
In the previous equation, $\mathcal{J}_{\alpha}(2t)$ is a Bessel function of the first kind of order $\alpha$.
Utilizing the above result in combination with Eq.~\eqref{eq:RealSpace}, the time-evolved operators become
\begin{equation}
        c_{x}(t)=\sum_{y}i^{(x-y)}\mathcal{J}_{x-y}(2t)c_{y}.
\end{equation}

Transport is determined by determining the late-time behavior of the density-density correlation function.
First, write the time-evolved density operator,
\begin{equation}
    \label{eq:density}
    n_{x}(t)=\sum_{\alpha,\beta}(i)^{\alpha-\beta}\mathcal{J}_{x-\alpha}(2t)\mathcal{J}_{x-\beta}(2t)c^{\dagger}_{\alpha}c_{\beta}.
\end{equation}
The correlation function from the main-text can be rewritten in the form,
\begin{equation}
    \label{eq:cor}
    C_{xy}(t) = \frac{1}{2^{N}}\text{tr}(n_{x}(t)n_{y}(0)) - \frac{1}{4}.
\end{equation}
By inserting Eq.~\eqref{eq:density} into the previous equation, we find,
\begin{align}
\begin{split}
    \label{eq:correlation}
    C_{x,y}(t) & = \frac{1}{2^{N}}\sum_{\alpha,\beta,\gamma,\delta}\left[i^{\alpha-\beta+\gamma-\delta}\mathcal{J}_{x-\alpha}(2t)\mathcal{J}_{x-\beta}(2t)\mathcal{J}_{y-\gamma}(0)\mathcal{J}_{y-\delta}(0)\right]\text{tr}(c^{\dagger}_{\alpha}c_{\beta}c^{\dagger}_{\gamma}c_{\delta}) - \frac{1}{4}\\
    & = \frac{1}{2^{N}}\sum_{\alpha,\beta}\left[i^{\alpha-\beta}\mathcal{J}_{x-\alpha}(2t)\mathcal{J}_{x-\beta}(2t)\right]\text{tr}(c^{\dagger}_{\alpha}c_{\beta}n_{y})-\frac{1}{4}.
    \end{split}
\end{align}
we utilized the relation $\mathcal{J}_{\alpha - \beta}(0)=\delta_{\alpha,\beta}$ when moving between lines.
In the previous equation, $C_{x,y}=0$ when $\alpha \neq \beta$, in which the sum breaks into two separate cases: (i) $\alpha \neq y$ and (ii) $\alpha = y$.
Finally, using the identity $\sum_{\alpha=-\infty}^{\infty}\mathcal{J}_{x-\alpha}^{2}(z)=1$, we evaluate the correlation function to be,
\begin{equation}
    \label{FreeResult}
    C_{x,y}(t) = \frac{1}{4}\mathcal{J}^{2}_{x-y}(2t).
\end{equation}
To compare our analytical result to simulations, we utilize the non-interacting structure of the model to simulate large systems to late time, enabling reliable extraction hydrodynamic behavior.
The following section focuses on efficiently implementing free-fermion numerics to compute the correlation function.

\subsection{Efficient Free-Fermion Simulations.}
\label{subseq:FreeFermSim}

The Hamiltonian in the many-body basis is stored as a $2^{N}\times 2^{N}$ matrix, severely limiting the possible system size. However, utilizing the non-interacting structure, we can simulate the dynamics in the single particle basis; the Hamiltonian is a $N \times N$ matrix. While the system size is no longer a barrier, here, because the model is Brownian, it must be diagonalized at each time step, severely restricting the simulation time. Moreover, for the system to remain uncorrelated in time, $dt$ must be small~(at least less than $0.1$).
We now outline how to compute $C_{x,y}(t)$ numerically on the single-particle basis.
We begin with the general Hamiltonian,
\begin{equation}
    H = \sum_{ij}c_{i}^{\dagger}M_{ij}c_{j}.
\end{equation}
Here, the matrix $M$ is a $N \times N$ real symmetric matrix that is diagonalized as $M=VDV^{T}$ where $V$ are the eigenvectors satisfying $VV^{T}=\mathbbm{1}$ and $D=\text{diag}(\lambda_{1},\ldots,\lambda_{N})$ is a diagonal matrix.
The Hamiltonian now becomes,
\begin{align}
    H &= \sum_{ij}\sum_{kl}c_{i}^{\dagger}V_{ik}\lambda_{k}V_{jk}c_{j}\nonumber \\
    &=\sum_{k}\lambda_{k}d_{k}^{\dagger}d_{k}.
\end{align}
We have defined the normal mode operators, $d_{k}=\sum_{j}V_{jk}c_{j}$ and $d^{\dagger}_{k}=\sum_{i}c_{i}^{\dagger}V_{ik}$ which inherit the standard fermionic anti-commutation relations.
We now define the vectors,
$\hat{d}^{\dagger}=[d_{1}^{\dagger},\ldots, d_{N}^{\dagger}]=\hat{c}^{\dagger}V$ and $\hat{d}  =[d_{1},\ldots, d_{N}]^{T}=V^{T}\hat{c}$.
From the Heisenberg evolution of the normal mode operators, we determine the original fermion operators evolve as,
\begin{align}
   \hat{c}^{\dagger}(t)=\hat{c}^{\dagger}\left(V e^{iDt}V^{T}\right)\nonumber \\
   \hat{c}(t)=\left(V e^{iDt}V^{T}\right)\hat{c}
\end{align}
The correlation function on the same site becomes,
\begin{widetext}
\begin{align}
    C_{x,x}(t) &= \frac{1}{2^{N}}\text{tr}\left(n_{x}(t)n_{x}\right) - \frac{1}{4}\nonumber \\
    &= \frac{1}{2^{N}}\left(\sum_{\sigma, \rho}\sum_{\mu, \nu}(V_{\sigma\rho}e^{i\lambda_{\rho}t}V_{\rho x}^{T})(V_{x \mu}e^{-i\lambda_{\mu}t}V_{\mu \nu}^{T})\text{tr}(c^{\dagger}_{\sigma}c_{\nu}c^{\dagger}_{x}c_{x}) \right)- \frac{1}{4}\nonumber \\
    &=\frac{1}{2^{N}}\left( \frac{2^{N}}{2}\sum_{\rho,\mu}V_{x \rho}V_{\rho x}^{T}V_{x\mu}V_{\mu x}^{T}e^{i\omega_{\rho\mu}t} + \frac{2^{N}}{4}\sum_{\rho,\sigma,\mu}V_{\sigma \rho}V_{\rho x}^{T}V_{x\mu}V_{\mu \sigma}^{T}e^{i\omega_{\rho\mu}t}\right)- \frac{1}{4}\nonumber \\
    & = \frac{1}{2}a_{x,x}(t)a^{*}_{x,x}(t)+\frac{1}{4}\sum_{\sigma,\sigma\neq x}a_{\sigma, x}(t)a^{*}_{x, \sigma}(t)- \frac{1}{4}.
\end{align}
\end{widetext}
Where above we have defined the time-dependent amplitudes, $a_{m,n}(t)=\sum_{k}V_{m,k}e^{i\lambda_{k}t}V^{T}_{k,n}$.
Moving from the first to the second line, we utilized that $\text{tr}(c^{\dagger}_{\sigma}c_{\nu}c^{\dagger}_{x}c_{x})=0$ if $\sigma \neq \nu$.
In the third line we consider the two cases: $\sigma = x$ and $\sigma \neq x$, where we used the relations $\text{tr}(n_{\sigma})=2^{N}/2$ and $\text{tr}(n_{\sigma}n_{x})=2^{N}/4$ and defined the energy difference, $\omega_{\rho \mu}=\lambda_{\rho}-\lambda_{\mu}$.
Efficient time evolution is performed as follows:
\begin{enumerate}[Step 1:]
    \item At each $dt$ draw $N-1$ random numbers from the Gaussian distribution with a mean $\mu = 0$ and standard deviation $\sigma = \left(\Gamma/dt\right)^{1/2}$.
    \item Diagonalize the $N \times N$ Hamiltonian.
    \item Compute the correlation function $C_{xx}(dt)$ from the eigenvalues and eigenvectors.
    \item Repeat the above steps for \textit{each} time-step.
    \item Repeat the entire procedure a designated number of disorder realizations and average.
\end{enumerate}

\section{Operator Equation of Motion.}
\label{appsec:app1}

After noise averaging, we now outline how to derive the effective Liouvillian that governs the operator dynamics. We consider the Brownian couplings, $\Gamma_{x,y}(t)$, which are time-dependent real Gaussian numbers with a mean and variance given by,
\begin{align}
\label{eq:Couplings}
\begin{split}
\mathbb{E}[\Gamma_{i,j}(t)] & = 0\\ \mathbb{E}[\Gamma_{i,j}(t)\Gamma_{l,m}(t^\prime)] & =\Gamma_{x,y} \delta_{i,l}\delta_{j,m}\delta(t-t^\prime).
\end{split}
\end{align}
Where the operation $\mathbb{E}[\cdot]$ denotes the Gaussian average couplings, $\Gamma_{x,y}$ sets the energy scale of the noise, and $\delta(t-t^\prime)$ implies that the couplings are correlated at a single instance in time.

We expand the flow of the Heisenberg operator up to order $dt^{2}$ because the variance is of order $dt^{-1}$. Specifically, consider the Heisenberg evolution of an operator $\mathcal{O}$,
\begin{align}
\begin{split}
\label{eq:Motion}
\mathcal{O}(t + dt) & = \mathcal{U}_{dt}\mathcal{O}(t)\mathcal{U}^{\dagger}_{dt}\\
    & = \mathcal{O}(t) + i[H_{t}, \mathcal{O}]dt + \mathcal{L}[\mathcal{O}]dt^{2}
    \end{split}
\end{align}
With the unitary operator, $\mathcal{U}_{dt}=e^{-iH_{t}dt}$ and the effective Lindblad, $\mathcal{L}[\mathcal{O}]=H^{\dagger}_{t}\mathcal{O}H_{t} - \frac{1}{2} \{H^{\dagger}_{t}H_{t},\mathcal{O} \}$.
We also emphasize that the above convention differs from standard transport theories in condensed matter that study the Green function, where the operator equation is put into a Schrodinger equation-like form.
Inserting the Hamiltonian into the operator equation, we get,
\begin{align}
    \mathcal{O}(t + dt) -\mathcal{O}(t) = \sum_{x,y}i\bigg[ J_{x,y}[c^{\dagger}_{x}c_{y}, \mathcal{O}] + \Gamma_{x,y}(t)[c^{\dagger}_{x}c_{y}, \mathcal{O}] \bigg]dt + \mathcal{L}_{x,y}[\mathcal{O}]dt^{2}
\end{align}
The Lindblad term breaks into terms with: $J_{x,y}J_{x,y}$, $J_{x,y}\Gamma_{x,y}(t)$, $\Gamma_{x,y}(t)\Gamma_{x,y}(t)$, moreover, only the first and final term survive noise averaging.
We now average over disorder realizations,
\begin{align}
    d\bar{\mathcal{O}_{t}} &= \sum_{x,y} i\bigg[ J_{x,y}[c^{\dagger}_{x}c_{y}, \bar{\mathcal{O}_{t}}] + \mathbb{E}[\Gamma_{x,y}(t)][c^{\dagger}_{x}c_{y}, d\bar{\mathcal{O}_{t}}] \bigg]dt +\bigg[ \mathbb{E}[\Gamma_{x,y}(t)\Gamma_{x,y}(t)]\mathcal{L}_{x,y}[\bar{\mathcal{O}_{t}}] + J_{x,y}J_{x,y}\mathcal{L}_{x,y}[\bar{\mathcal{O}_{t}}] \bigg]dt^{2}\nonumber \\
    &=\sum_{x,y} iJ_{x,y}[c^{\dagger}_{x}c_{y}, \bar{\mathcal{O}_{t}}]dt + \Gamma_{x,y} \mathcal{L}_{x,y}[\bar{\mathcal{O}_{t}}]dt + J_{x,y}J_{x,y}\mathcal{L}_{x,y}[\bar{\mathcal{O}_{t}}]dt^{2}.
\end{align}
Only keeping terms that are of order $dt$, leads to the noise-averaged operator equation,
\begin{align}
\label{eq:Average}
\frac{d\bar{\mathcal{O}_{t}}}{dt} =\sum_{x,y}\bigg[iJ_{x,y}[c^{\dagger}_{x}c_{y}, \bar{\mathcal{O}_{t}}]+\Gamma_{x,y} \mathcal{L}_{x,y}[\bar{\mathcal{O}_{t}}]\bigg]
\end{align}
Where we have defined $\mathbb{E}(d\mathcal{O})=d\bar{\mathcal{O}}$ and the Lindblad has the form, $\mathcal{L}_{x,y}[\bar{\mathcal{O}}]=L^{\dagger}_{x,y}\bar{\mathcal{O}}L_{x,y} - \frac{1}{2} \{L_{x,y}^{\dagger}L_{x,y},\bar{\mathcal{O}} \}$.
Generically, the above equation applies for long-range hopping, however, we consider either onsite or nearest-neighbor interactions.
Specifically, when the noise is coupled to nearest-neighbor hopping~($y=x+1$) then the jump operator is $L_{x,x+1}=c^{\dagger}_{x}c_{x+1}+c^{\dagger}_{x+1}c_{x}$, whereas, when the noise is coupled to the density $L_{x,x}=n_{x}$.
The most general operator equation of Eq.~\eqref{eq:Average} we consider is written as follows,
\begin{equation}
\label{eq:FullOpEq}
    \frac{d\bar{\mathcal{O}_{t}}}{dt} = \sum_{x} \bigg[iJ[L_{x,x+1}, \bar{\mathcal{O}_{t}}] -i\gamma x[L_{x,x},\bar{\mathcal{O}_{t}}] + \mathcal{V}\mathcal{L}_{x,x}[\bar{\mathcal{O}_{t}}] + \Gamma \mathcal{L}_{x,x+1}[\bar{\mathcal{O}_{t}}] \bigg].
\end{equation}
The above equation follows from setting the diagonal~(off-diagonal) elements of the static coupling matrix to $J_{x,x}=-\gamma x$~(we focus on a linear potential) and $J_{x,x+1}=J$, while the diagonal~(off-diagonal) elements of the noisy coupling matrix to $\Gamma_{x,x}=\mathcal{V}$ and $\Gamma_{x,x+1}=\Gamma$.

\section{Details in Calculating the Coefficient Differential Equation.}
In this section, we derive the differential equation for the operator amplitudes by inserting $n_{x}(t)$ into Eq.~\eqref{eq:Average}.

\subsection{Coherent Term from Static Hopping}
We begin with the time averaged operator equation of motion with only the static and noisy hopping contribution,
\begin{equation}
    \frac{d\bar{n}_{t}}{dt} =\sum_{x}\bigg[ iJ[L_{x, x+1}, \bar{n}_{t}] +\Gamma \mathcal{L}_{x,x+1}[\bar{n}_{t}]\bigg].
\end{equation}
In this section, we derive the term from the unitary dynamics,
\begin{align}
    [c^{\dagger}_{x}c_{x+1}, \bar{n}_{t}]&=\sum_{x}\sum_{mn}A_{m,n}\bigg[ c^{\dagger}_{x} [c_{x+1},c^{\dagger}_{m} ]c_{n} + [c^{\dagger}_{x}, c^{\dagger}_{m}]c_{x+1}c_{n} + c^{\dagger}_{m}c^{\dagger}_{x}[c_{x+1}, c_{n}] + c^{\dagger}_{m}[c^{\dagger}_{x},c_{n}]c_{x+1} \bigg]\nonumber \\
    &=\sum_{x}\sum_{mn}A_{m,n}\bigg[c^{\dagger}_{x} \left(2c_{x+1}c^{\dagger}_{m}-\delta_{x+1,m} \right)c_{n} + 2c^{\dagger}_{x}c^{\dagger}_{m}c_{x+1}c_{n} + 2c^{\dagger}_{m}c^{\dagger}_{x}c_{x+1}c_{n} + c^{\dagger}_{m} \left(2c_{x}c^{\dagger}_{n}-\delta_{x,n} \right)c_{x+1}\bigg]\nonumber \\
    &= \sum_{x}\sum_{m,n}A_{m,n}\bigg[ 2c^{\dagger}_{x}c_{x+1}c^{\dagger}_{m}c_{n} - \delta_{x+1,m}c^{\dagger}_{x}c_{n} + 2c^{\dagger}_{m}c^{\dagger}_{x}c_{n}c_{x+1} - \delta_{x,n}c^{\dagger}_{m}c_{x+1} \bigg]\nonumber \\
    &= \sum_{x}\sum_{m,n}A_{m,n}\bigg[ \delta_{x+1, m} c^{\dagger}_{x}c_{n} - \delta_{x, n}c^{\dagger}_{m}c_{x+1}\bigg].
\end{align}
In the first line, we utilized a commutator identity between four operators and then used that $[A, B]=2AB - \{A,B\}$ where we then applied the anti-commutator relation for fermion operators.
For the second line, the product of four fermion operators cancels by anti-commuting the creation operators and then expand in the third line.
Because the model is non-interacting the final result cannot have more than a product of two fermion operators, thus we simplify further using the anti-commutation relation $\{ c^{\dagger}_{m}, c_{x+1}\}$ and end with the final equation.
The second commutator is found by a similar calculation or by setting $x+1\rightarrow x$ and $x\rightarrow x+1$ in the above equation,
\begin{align}
    [c^{\dagger}_{x+1}c_{x}, \bar{n}_{t}]&= \sum_{x}\sum_{mn}A_{m,n}\bigg[ \delta_{x, m} c^{\dagger}_{x+1}c_{n} - \delta_{x+1, n}c^{\dagger}_{m}c_{x}\bigg].
\end{align}
Moreover, we find the final equation after resolving the delta functions is,
\begin{equation}
    iJ\sum_{x}[L_{x, x+1}, \bar{n}_{t}] = iJ\sum_{m,n}A_{m,n}\bigg[ c^{\dagger}_{m-1}c_{n} - c^{\dagger}_{m}c_{n+1} + c^{\dagger}_{m+1}c_{n} - c^{\dagger}_{m}c_{n-1} \bigg]
\end{equation}
Next, we want to shift the indices so that each operator is equivalently $c^{\dagger}_{m}c_{n}$ which allow us to equate sides and get a differential equation for $A_{m,n}$.
\begin{align}
    A_{m,n}c^{\dagger}_{m-1}c_{n} &\rightarrow A_{m+1,n}c^{\dagger}_{m}c_{n},\quad A_{m,n}c^{\dagger}_{m}c_{n+1} \rightarrow A_{m,n-1}c^{\dagger}_{m}c_{n}\nonumber \\
    A_{m,n}c^{\dagger}_{m+1}c_{n} &\rightarrow A_{m-1,n}c^{\dagger}_{m}c_{n},\quad A_{m,n}c^{\dagger}_{m}c_{n-1} \rightarrow A_{m,n+1}c^{\dagger}_{m}c_{n}
\end{align}
The final equation due to unitary dynamics is,
\begin{equation}
    iJ\sum_{x}[L_{x, x+1}, \bar{n}_{t}] = iJ\sum_{mn}\bigg[A_{m+1,n} - A_{m,n-1} + A_{m-1,n} - A_{m,n+1}\bigg]c^{\dagger}_{m}c_{n}
\end{equation}
In the next section, we derive the contribution from the effective Lindblad term.

\subsection{Incoherent Term from Noisy Hopping.}

In this section, we derive the incoherent piece that arises from the variance of the couplings,
\begin{equation}
    \sum_{x}\mathcal{L}_{x,x}\big[ \bar{n}_{t} \big] = \sum_{x}\sum_{m,n}A_{m,n}\bigg[ L^{\dagger}_{x,x+1}c^{\dagger}_{m}c_{n}L_{x,x+1} - \frac{1}{2} \{ L^{\dagger}_{x,x+1}L_{x,x+1}, c^{\dagger}_{m}c_{n} \} \bigg].
\end{equation}
where $L_{x,x+1}=c^{\dagger}_{x}c_{x+1}+c^{\dagger}_{x+1}c_{x}$.
After some algebra and using fermion identities, we get the final equation,
\begin{align}
\frac{d\bar{n}_{t}}{dt} &= \sum_{mn}\bigg[-2A_{m,n}c^{\dagger}_{m}c_{n} + A_{m,n}\delta_{m,n-1}c^{\dagger}_{n}c_{n-1}+ A_{m,n}\delta_{m,n+1}c^{\dagger}_{n}c_{n+1} +A_{m,n}\delta_{m,n}\left(n_{m+1}+n_{m-1} \right) \bigg]
\end{align}
Next, we want to shift the indices so that each operator is aligned with $c^{\dagger}_{m}c_{n}$ which allow us to equate sides and get a differential equation for $A_{m,n}$.
\begin{align}
    A_{m,n}\delta_{m,n}c^{\dagger}_{m-1}c_{m-1} &\rightarrow A_{m+1,n+1}\delta_{m,n}c^{\dagger}_{m}c_{n},\quad A_{m,n}\delta_{m,n}c^{\dagger}_{m+1}c_{m+1} \rightarrow A_{m-1,n-1}\delta_{m,n}c^{\dagger}_{m}c_{n}\nonumber \\
    A_{m,n}\delta_{m,n+1}c^{\dagger}_{n}c_{n+1} &\rightarrow A_{m+1,n-1}\delta_{m+1,n}c^{\dagger}_{m}c_{n},\quad A_{m,n}\delta_{m,n-1}c^{\dagger}_{n}c_{n-1} \rightarrow A_{m-1,n+1}\delta_{m-1,n}c^{\dagger}_{m}c_{n}
\end{align}
The operator equation of motion becomes,
\begin{align}
\frac{d\bar{n}_{t}}{dt} &= \sum_{m,n}\bigg[-2A_{m,n} + \delta_{m+1,n}A_{m+1,n-1} + \delta_{m-1,n}A_{m-1,n+1} + \delta_{m,n}A_{m+1,n+1}+ \delta_{m,n}A_{m-1,n-1}\bigg]c^{\dagger}_{m}c_{n}.
\end{align}
Moreover, we find that the contribution from the incoherent piece is as follows,
\begin{equation}
    \frac{dA_{m,n}}{dt} = \bigg[-2A_{m,n} + \delta_{m+1,n}A_{m+1,n-1} + \delta_{m-1,n}A_{m-1,n+1} + \delta_{m,n}A_{m+1,n+1} + \delta_{m,n}A_{m-1,n-1}\bigg]
\end{equation}
In the following section we will perform the mapping to the fictitious lattice defined by the operator length and center-of-mass coordinates.

\subsection{Operator Equation of Motion.}
Combining, the equation from the previous two sections, we arrive at the full operator equation,
\begin{align}
&\partial_{t} A_{m,n}  = iJ\bigg[A_{m+1,n}-A_{m,n+1}+A_{m-1,n}-A_{m,n-1} \bigg]\nonumber \\ &+ \Gamma \bigg[-2A_{m,n}+\delta_{m+1, n}A_{m+1, n-1}+ \delta_{m-1, n}A_{m-1, n+1}+ \delta_{m,n}A_{m-1,n-1}+\delta_{m,n}A_{m+1,n+1}\bigg].
\end{align}
When a static potential is included in the coherent dynamics, the additional term is $i\big[V_{m}-V_{n}\big]A_{m,n}$ which has no affect in the strong noise limit.

\subsection{Mapping to $\ell$ and $\mathcal{R}$.}
Spatial translation invariance is partially restored in a different coordinate system defined by the variables $\ell=n-m$ and $\mathcal{R}=n+m$.
Here $\ell$ represents the length of the operator and $\mathcal{R}$ the center-of-mass of the operator.
The coefficients under this mapping become,
\begin{align}
    A_{m+1,n-1} \overset{(n-1-m-1=\ell-2, n-1+m+1=\mathcal{R})}{\longrightarrow} A_{\ell-2, \mathcal{R}},\quad &A_{m-1,n+1} \overset{(n+1-m+1=\ell+2, n+1+m-1=\mathcal{R})}{\longrightarrow} A_{\ell+2, \mathcal{R}}\nonumber \\
    A_{m+1,n+1} \overset{(n+1-m-1=\ell, n+1+m+1=\mathcal{R}+2)}{\longrightarrow} A_{\ell, \mathcal{R}+2},\quad &A_{m-1,n-1} \overset{(n-1-m+1=\ell, n-1+m-1=\mathcal{R}-2)}{\longrightarrow} A_{\ell, \mathcal{R}-2}\nonumber \\
    \delta_{m+1,n} \overset{(\ell=1)}{\longrightarrow} \delta_{\ell, 1},\quad &\delta_{m-1,n} \overset{\ell=-1)}{\longrightarrow} \delta_{\ell,-1}
\end{align}
Applying this transformation to the equation above,
\begin{align}
    \partial_{t} A_{\ell,\mathcal{R}} &= iJ\bigg[ A_{\ell-1, \mathcal{R}+1}-A_{\ell+1, \mathcal{R}+1}+A_{\ell+1, \mathcal{R}-1}-A_{\ell-1, \mathcal{R}-1} \bigg]\nonumber\\ &+ \Gamma\bigg[ -2A_{\ell,\mathcal{R}}+\delta_{\ell,1}A_{\ell-2, \mathcal{R}}+\delta_{\ell,-1}A_{\ell+2,\mathcal{R}} + \delta_{\ell,0}\left(A_{\ell,\mathcal{R}+2} +A_{\ell,\mathcal{R}-2} \right) \bigg]
\end{align}
Now the above equation is simplified further by looking at the terms $\delta_{\ell,-1}A_{\ell+2,\mathcal{R}}$ and $\delta_{\ell,1}A_{\ell-2, \mathcal{R}}$ which are nonzero if $\ell=\pm 1$ with the coefficients $A_{\mp 1}$.
A similar argument can be made for $\delta_{\ell,0}\left(A_{\ell,\mathcal{R}+2} +A_{\ell,\mathcal{R}-2} \right)$.
Moreover, the final equation is,
\begin{align}
    \partial_{t} A_{\ell,\mathcal{R}} &= iJ\bigg[ A_{\ell-1, \mathcal{R}+1}-A_{\ell+1, \mathcal{R}+1}+A_{\ell+1, \mathcal{R}-1}-A_{\ell-1, \mathcal{R}-1} \bigg]\nonumber \\ &+ \Gamma\bigg[ -2A_{\ell,\mathcal{R}}+\delta_{\ell,1}A_{-1, \mathcal{R}}+\delta_{\ell,-1}A_{1,\mathcal{R}} + \delta_{\ell,0}\left(A_{0,\mathcal{R}+2} +A_{0,\mathcal{R}-2} \right) \bigg]
\end{align}
Now because the delta functions are independent of $\mathcal{R}$ we can perform a discrete Fourier transformation in this coordinate, namely, $A_{\ell,\mathcal{R}}=\sum e^{ik\mathcal{R}}A_{\ell,k}$.
The Fourier transformed equation becomes,
\begin{align}
    \partial_{t} A_{\ell,k} &= 2J\sin(k) \bigg[A_{\ell+1, k} - A_{\ell-1,k} \bigg] + \Gamma\bigg[-2A_{\ell,k}+\delta_{\ell,1}A_{-1,k}+\delta_{\ell,-1}A_{1,k}+2\cos(2k)\delta_{\ell,0}A_{0,k}\bigg]
\end{align}
In the next section, we derive the contribution from a noisy potential, specifically, $\Gamma_{x,x} \neq 0$.

\subsection{Incoherent Term from Noisy Potential.}

In this section, we derive the incoherent piece that arises from the variance of the couplings,
\begin{equation}
    \sum_{x}\mathcal{L}_{x}\big[ \bar{n}_{t} \big] = \sum_{x}\sum_{m,n}A_{m,n}\bigg[ n_{x}c^{\dagger}_{m}c_{n}n_{x} - \frac{1}{2} \{ n_{x}, c^{\dagger}_{m}c_{n} \} \bigg].
\end{equation}
As in the previous section, the final result must be a product of two fermionic operators.
We first calculate the second term in the above equation,
\begin{align}
    \{n_{x},c^{\dagger}_{m}c_{n}\} &= c^{\dagger}_{x}c_{x}c^{\dagger}_{m}c_{n} + c^{\dagger}_{m}c_{n}c^{\dagger}_{x}c_{x}\nonumber \\
    &= c^{\dagger}_{x} \left( \delta_{m,x}-c^{\dagger}_{m}c_{x} \right)c_{n} + c^{\dagger}_{m}\left(\delta_{x,n}-c^{\dagger}_{x}c_{n} \right)c_{x}\nonumber \\
    &= \delta_{m,x}c^{\dagger}_{x}c_{n} +c^{\dagger}_{m}c^{\dagger}_{x}c_{x}c_{n} + c^{\dagger}_{m}c^{\dagger}_{x}c_{x}c_{n} + \delta_{n,x}c^{\dagger}_{m}c_{x}\nonumber \\
    &= 2c^{\dagger}_{m}c^{\dagger}_{x}c_{x}c_{n} + \delta_{m,x}c^{\dagger}_{x}c_{n} + \delta_{n,x}c^{\dagger}_{m}c_{x}.
\end{align}
We put the fermion operators in normal order to simplify all these expressions.
Therefore, in the first equation above, we use the anti-commutator to move the creation operators to the left and then do algebra on the subsequent lines.
Combing everything, we have that,
\begin{equation}
    -\frac{1}{2}\{n_{x},\bar{n}_{t} \} = -\sum_{x}\sum_{m,n}A_{m,n} c^{\dagger}_{m}c^{\dagger}_{x}c_{x}c_{n} - \sum_{mn}A_{m,n}c^{\dagger}_{m}c_{n}.
\end{equation}
We resolved the delta functions in the second term to eliminate the sum over x.
Note that if $A_{m,n}=\delta_{x,m}\delta_{x,n}$ we get that the above equation is $-n_{x}$ which we expect from $-\frac{1}{2}\{ n_{x}, n_{x} \}=-n_{x}$.
Another check is if $A_{m,n}=\delta_{y,m}\delta_{y,n}$ then we find $-n_{y}n_{x}$ as expected.
We now compute the first term, which is a product of six fermion operators,
\begin{align}
    c^{\dagger}_{x}c_{x}c^{\dagger}_{m}c_{n}c^{\dagger}_{x}c_{x} &= c^{\dagger}_{x}\left(\delta_{mx} - c^{\dagger}_{m}c_{x} \right)c_{n}c^{\dagger}_{x}c_{x}\nonumber \\
    &= \delta_{m,x}c^{\dagger}_{x}c_{n}c^{\dagger}_{x}c_{x} - c^{\dagger}_{x}c^{\dagger}_{m}c_{x}c_{n}c^{\dagger}_{x}c_{x}\nonumber \\
    &= \delta_{m,x}c^{\dagger}_{x}c_{n}c^{\dagger}_{x}c_{x} - \delta_{n,x}c^{\dagger}_{x}c^{\dagger}_{m}c_{x}c_{x} + c^{\dagger}_{x}c^{\dagger}_{m}c_{x}c^{\dagger}_{x}c_{n}c_{x}\nonumber \\
    &= \delta_{m,x}c^{\dagger}_{x}c_{n}c^{\dagger}_{x}c_{x} + c^{\dagger}_{x}c^{\dagger}_{m}c_{n}c_{x}
\end{align}
In total we have that,
\begin{equation}
    \sum_{x}\sum_{m,n}A_{m,n}c^{\dagger}_{x}c_{x}c^{\dagger}_{m}c_{n}c^{\dagger}_{x}c_{x} = \sum_{x}\sum_{mn}A_{m,n}\bigg[\delta_{m,x}c^{\dagger}_{x}c_{n}c^{\dagger}_{x}c_{x} + c^{\dagger}_{x}c^{\dagger}_{m}c_{n}c_{x} \bigg]
\end{equation}
We now combine the previous two results,
\begin{align}
    \sum_{x}\mathcal{L}_{x}\big[ \bar{n}_{t} \big] &= \sum_{x}\sum_{m,n}A_{m,n}c^{\dagger}_{x}c_{x}c^{\dagger}_{m}c_{n}c^{\dagger}_{x}c_{x} -\frac{1}{2}A_{m,n}\{n_{x}, c^{\dagger}_{m}c_{n} \}\nonumber \\
    &= \sum_{x}\sum_{m,n}A_{m,n}\bigg[\delta_{m,x}c^{\dagger}_{x}c_{n}c^{\dagger}_{x}c_{x} + c^{\dagger}_{x}c^{\dagger}_{m}c_{n}c_{x} \bigg] -\sum_{x}\sum_{m,n}A_{m,n} c^{\dagger}_{m}c^{\dagger}_{x}c_{x}c_{n} - \sum_{m,n}A_{m,n}c^{\dagger}_{m}c_{n}\nonumber \\
    &= \sum_{x}\sum_{m,n}A_{m,n}\delta_{m,x}c^{\dagger}_{x}c_{n}c^{\dagger}_{x}c_{x} - \sum_{m,n}A_{m,n}c^{\dagger}_{m}c_{n}\nonumber \\
    &= \sum_{m,n}A_{m,n}c^{\dagger}_{m}c_{n}c^{\dagger}_{m}c_{m} - \sum_{m,n}A_{m,n}c^{\dagger}_{m}c_{n}
\end{align}
There are a few checks we can perform on the above equation. First, if $A_{m,n}=\delta_{x,m}\delta_{x,n}$ or $A_{m,n}=\delta_{y,m}\delta_{y,n}$ the above result vanishes.
When the Lindblad acts on a number operator, i.e., $\mathcal{L}[c^{\dagger}_{\alpha}c_{\alpha}]$, it annihilates it.
The final simplification is the normal ordering of the first term above,
\begin{equation}
    \sum_{x}\mathcal{L}_{x}\big[ \bar{n}_{t} \big] = \sum_{m,n}A_{m,n} \left(\delta_{n,m}c^{\dagger}_{m}c_{m}- c^{\dagger}_{m}c_{n}\right)
\end{equation}
Notice above that if the operator is on the same site, i.e., $m=n$, then we get zero, and if $m\neq n$, we get $-\bar{n}_{t}$.
Thus we can rewrite the above equation as,
\begin{equation}
    \sum_{x}\mathcal{L}_{x}\big[ \bar{n}_{t} \big] = \sum_{mn}A_{m,n} \left(\delta_{nm}- 1\right)c^{\dagger}_{m}c_{n}
\end{equation}
Moreover, the final differential equation becomes,
\begin{equation}
    \frac{d\bar{n}_{t}}{dt} = \mathcal{V}\left(\delta_{nm}- 1\right)A_{m,n}
\end{equation}

\subsection{Mapping to $\ell$ and $\mathcal{R}$}
In this section, we rewrite the final equation of the previous section in a different coordinate system defined by the variables $\ell=n-m$ and $\mathcal{R}=n+m$.
Here $\ell$ represents the length of the operator, and $\mathcal{R}$ is the center of mass of the operator.
The coefficients under this mapping become,
\begin{align}
    A_{m+1,n} \overset{(n-m-1=\ell-1, n+m+1=\mathcal{R}+1)}{\longrightarrow} A_{\ell-1, \mathcal{R}+1},\quad &A_{m,n-1} \overset{(n-1-m=\ell-1, n-1+m=\mathcal{R}-1)}{\longrightarrow} A_{\ell-1, \mathcal{R}-1}\nonumber \\
    A_{m-1,n} \overset{(n-m+1=\ell+1, n+m-1=\mathcal{R}-1)}{\longrightarrow} A_{\ell+1, \mathcal{R}-1},\quad &A_{m,n+1} \overset{(n+1-m=\ell+1, n+1-m=\mathcal{R}+1)}{\longrightarrow} A_{\ell+1, \mathcal{R}+1}\nonumber \\
    A_{m,n} \overset{(n-m=\ell, n+m=\mathcal{R})}{\longrightarrow} A_{\ell, \mathcal{R}},\quad &\delta_{m,n}\overset{(n-m=0)}{\longrightarrow} \delta_{\ell,0}.
\end{align}
Moreover, the differential equation becomes,
\begin{equation}
 \frac{dA_{\ell,\mathcal{R}}}{dt} = \mathcal{V} A_{\ell,\mathcal{R}}\left(\delta_{\ell,0}- 1\right)
\end{equation}
The advantage of moving into this coordinate system is that the delta function is independent of $\mathcal{R}$; thus, we can perform a discrete Fourier transformation $A_{\ell, \mathcal{R}}=\sum_{k}e^{ik\mathcal{R}}A_{\ell,k}$ where $k$ is the momentum of the Brillouin zone.
The Fourier transformed equation becomes,
\begin{equation}
    \frac{dA_{\ell,k}}{dt} = \mathcal{V} A_{\ell,k}\left(\delta_{\ell,0}- 1\right).
\end{equation}
With all these pieces in hand, the differential equation from Eq.~\eqref{eq:FullOpEq} takes the form,
\begin{equation}
    \label{eq:CoeffDiffEq}
    \frac{dA_{\ell}}{dt} = t_{k} \big[A_{\ell +1} - A_{\ell - 1} \big] + \Gamma\bigg[-2A_{\ell,k}+\delta_{\ell,1}A_{-1,k}+\delta_{\ell,-1}A_{1,k}+2\cos(2k)\delta_{\ell,0}A_{0,k}\bigg] +i\gamma \ell A_{\ell} +\mathcal{V}\big[\delta_{\ell, 0} - 1\big]A_{\ell}.
\end{equation}
The following sections will focus on various limits of the above equation and solve for the diffusion constant exactly in the small momentum limit.

\section{Diffusion Constant for Static Hopping with Bond Noise.}

\begin{figure}[t]
    \includegraphics[width=\textwidth]{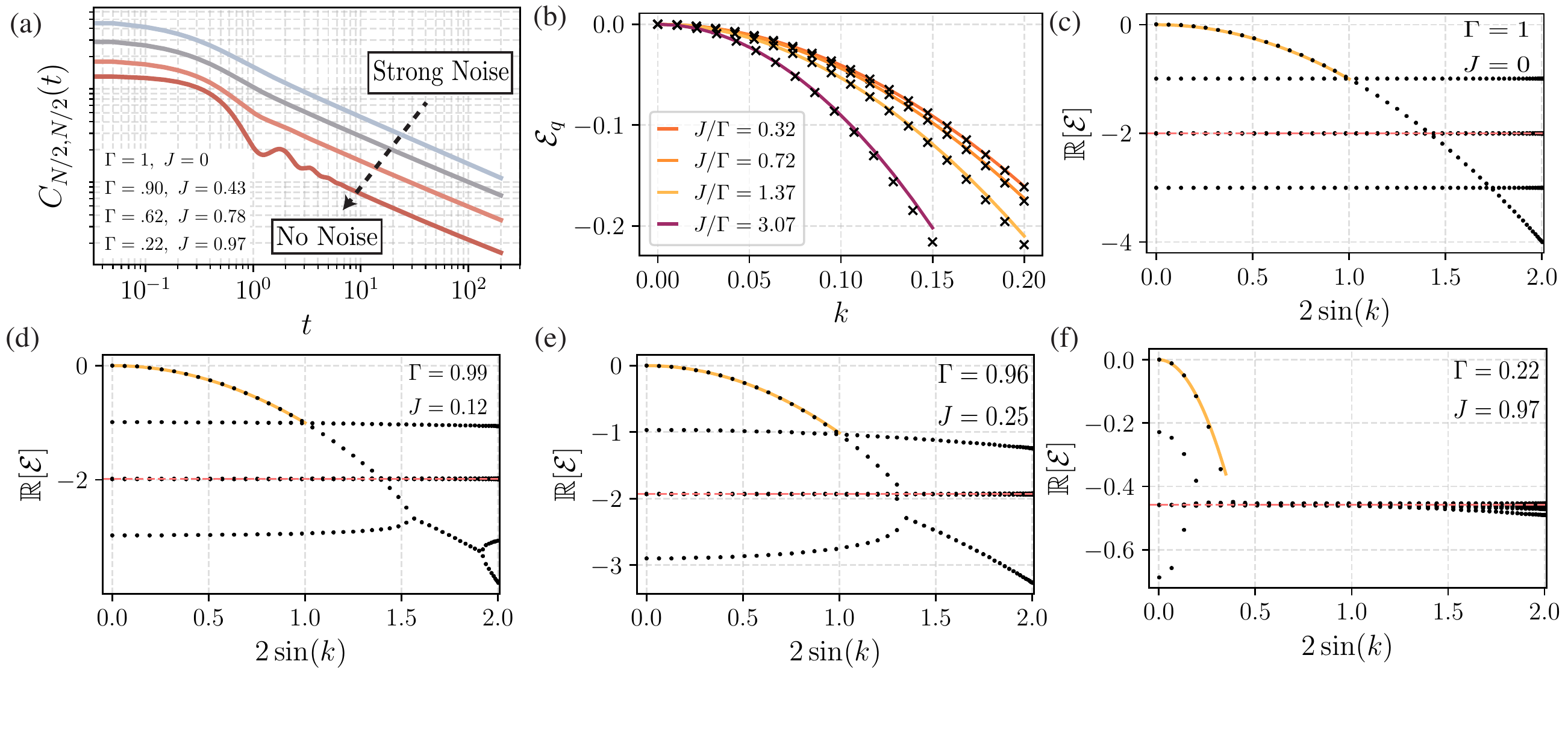}
    \caption{\textit{Static and Noisy Hopping Characteristics.}
    (a) Noisy-hopping operator dynamics.
    (b) Bound state energy for the case of static and noisy hopping, i.e., $\mathcal{V}=0$ and $\gamma = 0$. The solid curves are the analytical result, Eq.~\eqref{eq:FiniteNoiseEnergyApp} in the small momentum limit, while the black crosses are from diagonalizing the eigenvalue equations, Eq.~\eqref{eq:FiniteNoiseLimit}.
    (c)-(f) Real part of the eigenvalue spectrum of Eq.~\eqref{eq:FiniteNoiseLimit} for different noisy and static hopping strengths. Regardless of the noise strength, there is always a bound state at small momentum. As momentum increases, the bound state vanishes in a continuum of bound states at $-2\Gamma$ represented by the red line. Increasing $k$ further reveals that the bound state reemerges at an exceptional point. Before the exceptional point, the cubic equation admits two physical solutions with real energies, which then coalesce and become a conjugate pair of complex energies. Interestingly, near the strong drive limit there is a reentrant-like point~[see (d)] where the degeneracy is re-broken, and the purely real energies emerge once more.
    Parameters: (a) $N=400$, $dt=0.05$, $\ell_{\text{max}}=N/2$, (b) $N=400$, and (c)-(f) $N=400$, $\ell_{\text{max}}=N/2$.
    }
    \label{fig:BoundStateEnScale}
\end{figure}

In this section, we give an explicit solution to the bound state energy in the case of finite noise with no noisy or static potential, i.e., $\mathcal{V}=0$ and $\gamma=0$ in Eq.~\eqref{eq:CoeffDiffEq}.
We begin with the eigenvalue equation,
\begin{equation}
\label{eq:FiniteNoiseLimit}
    \mathcal{E}_{q}A_{\ell} = t_{k}\bigg[A_{\ell+1} - A_{\ell-1} \bigg] + \Gamma\bigg[-2A_{\ell}+\delta_{\ell,1}A_{-1}+\delta_{\ell,-1}A_{1}+2\cos(2k)\delta_{\ell,0}A_{0}\bigg].
\end{equation}
We have dropped the subscript $k$ since we are solving these for a fixed momentum and $t_{k}=2J\sin(k)$ is the effective hopping strength.
For finite noise, the above equation is non-hermitian and effectively describes the dynamics of a one-dimensional hopping model.
The above equation splits into four distinct eigenvalue equations,
\begin{align}
\label{eq:FiniteNoiseEquations}
        \mathcal{E}_{q}A_{0} &= t_{k}\left(A_{1}-A_{-1} \right) -4\Gamma \sin^{2}(k)A_{0},\quad \mathcal{E}_{q}A_{1} = t_{k}\left(A_{2}-A_{0} \right) -2\Gamma A_{1} + \Gamma A_{-1}\nonumber \\
        \mathcal{E}_{q}A_{-1} &= t_{k}\left(A_{0}-A_{-2} \right) -2\Gamma A_{-1} + \Gamma A_{1},\quad \mathcal{E}_{q}A_{\ell} = t_{k}\left(A_{\ell+1}-A_{\ell-1} \right) -2\Gamma A_{\ell},
\end{align}
The final equation describes the bulk dynamics and sets the energy of the problem, while the first three equations are boundary conditions that constrain the allowed values that $q$ can take.
For $J=0$, the above equations decouple. The eigenvalue equation for $A_{0}$ maps to the standard diffusion equation, and the entire spectrum forms horizontal lines at $\mathcal{E}=-1,-2,-3$ where the fourth value is the diffusive mode $-4k^{2}$ visualized in Fig.~\eqref{fig:BoundStateEnScale}(c).
On the other hand, for $\Gamma = 0$, the model is free with ballistic transport.
We now attempt to solve for the diffusion constant for generic noise strength, $\Gamma$.
First, we solve the bulk equation with the solution, $A_{\ell}=Ae^{q\ell} + Be^{-q\ell + i\pi\ell}$,
\begin{equation}
\label{eq:FiniteNoiseEnergy}
    \mathcal{E}_{q} = 4J\sin(k)\sinh(q) - 2\Gamma
\end{equation}
Generically, $q$ depends on momentum $k$ which we determine through the equation $A_{0}$.
First, if $A_{1}$ and $A_{-1}$ are known, then the entire set of coefficients is determined through the above equations; thus we use the ansatz,
\begin{equation}
\label{eq:FiniteNoiseAnsatz}
    A_{\ell} = \begin{cases}
    A_{-1}e^{q(1+\ell)} & \text{if}\ \ell \leq -1\\
    \\
    -A_{1}e^{q(1-\ell)+i\pi\ell}& \text{if}\ \ell \geq 1
    \end{cases}
\end{equation}
The three remaining equations solve for the variables $A_{1}/A_{-1}$, $A_{0}/A_{-1}$, and $q$.
The middle two equations solve for the ratios, and the equation for $A_{0}$ restricts the values $q$ takes.
First, rewrite the middle two equations,
\begin{align}
    \bigg[ \mathcal{E}_{q} + 2Je^{-q}\sin(k) + 2\Gamma \bigg]\frac{A_{1}}{A_{-1}} + 2J\sin(k)\frac{A_{0}}{A_{-1}} = \Gamma\nonumber \\
    \bigg[ \mathcal{E}_{q} + 2Je^{-q}\sin(k) + 2\Gamma \bigg] - 2J\sin(k)\frac{A_{0}}{A_{-1}} = \Gamma\frac{A_{1}}{A_{-1}}
\end{align}
Solving these equations yields,
\begin{equation}
\label{eq:RatioConstraint}
    \frac{A_{1}}{A_{-1}} = -1,\quad \frac{A_{0}}{A_{-1}} = e^{q} + \frac{\Gamma }{2J\sin(k)}
\end{equation}
Here $A_{1}$ and $A_{-1}$ are determined through normalizing $A_{\ell}$. The value of $A_{0}$ is determined through the second equation.
The ratios above, combined with the final eigenvalue equation for $A_{0}$ leads to the constraint,
\begin{equation}
\label{eq:QconstraintEquation}
    \bigg[\mathcal{E}_{q}+4\Gamma \sin^{2}(k) \bigg] \bigg[ e^{q} + \frac{\Gamma }{2J\sin(k)}\bigg] = -4J\sin(k)
\end{equation}
The above equation is a cubic equation for $q$, which can be solved exactly.
At small momentum, all three solutions to the cubic equation are physical and admit purely real energies; one corresponds to the largest real eigenvalue, the bound state energy. We solve the cubic equation and expand the bound state energy around $k=0$ giving $q$,
\begin{equation}
    e^{q}= -\frac{\Gamma}{Jk} - \bigg[\frac{J}{3 \Gamma}+\frac{11 \Gamma }{6 J} \bigg]k
\end{equation}
Inserting the above solution into the equation for the energy and keeping the lowest order in $k$, we find,
\begin{align}
\label{eq:FiniteNoiseEnergyApp}
    \mathcal{E}_{q} &= 2kJ\left( e^{q}-e^{-q}\right) - 2\Gamma \approx -\bigg[\frac{8}{3}\frac{J^{2}}{\Gamma} -4\Gamma \bigg]k^{2}\nonumber \\
    \mathcal{D} &= \bigg[\frac{8}{3}\frac{J^{2}}{\Gamma} -4\Gamma \bigg]
\end{align}
Moreover, regardless of the strength of the noisy hopping, the late-time hydrodynamic tail remains diffusive.
In Fig.~\ref{fig:BoundStateEnScale}(c)-(f), we diagonalize Eq.~\eqref{eq:FiniteNoiseEquations} for different noise and static strengths and plot the real part of the spectrum.
For all strengths, the diffusive mode is present at small $k$, which we fit with the analytical result Eq.~\eqref{eq:FiniteNoiseEnergyApp}~[see yellow curve].
As momentum increases, the bound state undergoes a phase transition by entering a continuum of scattering states at $-2\Gamma$, indicated by the red line in Fig.~\ref{fig:BoundStateEnScale}(c)-(f). Interestingly as $k$ increases further, the bound state reemerges at an exceptional like point. At this point, the two physical solutions to the cubic equation, which give purely real energies, collide and become a pair of conjugate energies. Interestingly, there is a higher-order exceptional-like point near the strong noise limit where the solutions transition back to purely real non-degenerate energies; see Fig.~\ref{fig:BoundStateEnScale}(d). We also emphasize that an alternative solution to the above eigenvalue equations is to assume $A_{0}=0$, in which case you can determine analytically the form of the branch beginning at $-\Gamma$ in Fig.~\ref{fig:BoundStateEnScale}(c)-(f).

\section{Diffusion Constant for Static Hopping and Onsite Noise.}

\begin{figure}[t]
    \includegraphics[width=\textwidth]{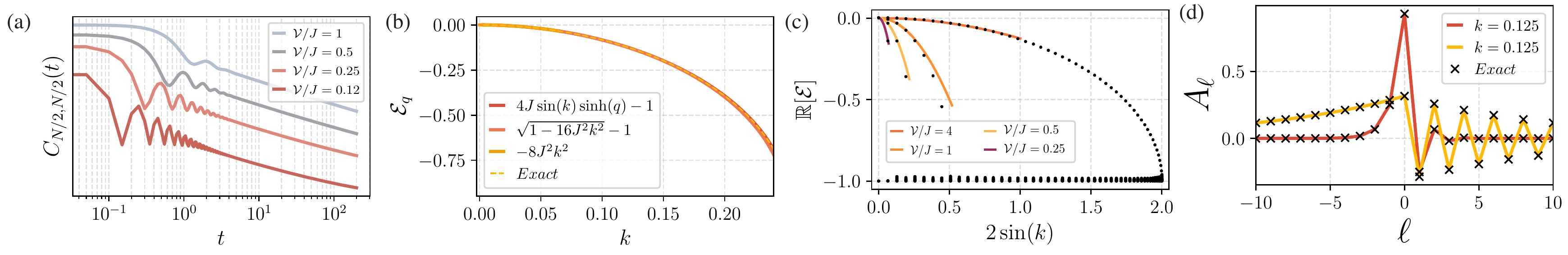}
    \caption{\textit{Static Hopping and Noisy Potential Characteristics.}
    (a) Operator dynamics for static hopping with a noisy potential.
    (b) The bound state energy for the noisy potential with static hopping. The dotted orange line is from diagonalizing the eigenvalue equation with the other curves representing different stages in approximating the low momentum scaling.
    (c) Real part eigenvalue spectrum for the noisy potential, where there is a single bound state corresponding to the diffusive mode with energy given by Eq.~\eqref{eq:AppPotBSE}.
    (d) The coefficients $A_{\ell}$ from Eq.~\eqref{eq:AppSHNPAnsatz} for different $k$ compared to diagonalizing of Eq.~\eqref{eq:AppSHNP}.
    Parameters: (a) $N=400$, $dt=0.05$, $\ell_{\text{max}}=N/2$, (b) $N=4000$, $\mathcal{V}/J=1$, and (c) $N=4000$ (d) $N=4000$, $\mathcal{V}/J=1$.
    }
    \label{fig:BoundStateEnScalePot}
\end{figure}

In this section, we solve Eq.~\eqref{eq:CoeffDiffEq} with $\Gamma=\gamma=0$.
We begin with the eigenvalue equation,
\begin{equation}
\label{eq:AppSHNP}
    \mathcal{E}_{q}A_{\ell,k} = t_{k}\left(A_{\ell+1, k}-A_{\ell-1, k} \right) + \mathcal{V}\left( \delta_{\ell,0}-1\right)A_{\ell,k}
\end{equation}
We now split the above equation into two different equations,
\begin{align}
        \mathcal{E}_{q}A_{0,k} &= t_{k}\left(A_{1, k}-A_{-1, k} \right)\nonumber \\
        \mathcal{E}_{q}A_{\ell,k} &= t_{k}\left(A_{\ell+1, k}-A_{\ell-1, k} \right) -\mathcal{V} A_{\ell,k},
\end{align}
where $\ell$ represents the entire line without zero.
Using the second equation, we can determine the energy by assuming the ansatz, $A_{\ell}=Ae^{q\ell} + Be^{-q\ell + i\pi\ell}$ leading to the relation,
\begin{align}
   (\mathcal{V}+\mathcal{E}_{q})\left(Ae^{q\ell} + Be^{-q\ell + i\pi\ell} \right) &= 2J\sin(k) \bigg[Ae^{q\ell}(e^{-q}-e^{q}) + Be^{-q\ell + i\pi\ell}(e^{q+i\pi}-e^{-q+i\pi}) \bigg]\nonumber \\
   &= 4J\sin(k)\sinh(q)\left( Ae^{q\ell} + Be^{-q\ell + i\pi\ell}\right).
\end{align}
Moreover, the energy is determined as
\begin{equation}
    \mathcal{E}_{q} = 4J\sin(k)\sinh(q) - \mathcal{V}
\end{equation}
The next step is to determine the values $q$ can take.
First, we write the solution to the second equation in both regions,
\begin{equation}
    A_{\ell} = \begin{cases}
    Ae^{q\ell} + Be^{-q\ell + i\pi\ell}& \text{if}\ \ell < 0\\
    Ce^{q\ell} + De^{-q\ell + i\pi\ell}& \text{if}\ \ell > 0
    \end{cases}
\end{equation}
In the region $\ell \leq 0$ the second term is unbounded as $\ell \rightarrow -\infty$ thereby we set $B=0$.
Similarly, for $\ell \geq 0$ the first term is unbounded as $\ell \rightarrow \infty$ and we therefore also set $C=0$.
A further constraint is made at $\ell=0$ which sets $A=D$ and the general solution becomes,
\begin{equation}
\label{eq:AppSHNPAnsatz}
    A_{\ell} = \begin{cases}
    Ae^{q\ell} & \text{if}\ \ell \leq 0\\
    Ae^{-q\ell + i\pi\ell}& \text{if}\ \ell \geq 0.
    \end{cases}
\end{equation}
With this solution, we can determine $A_{1}$ and $A_{-1}$ and inset them into the first equation above for $A_{0}$ which leads to the constraint on $q$,
\begin{align}
\mathcal{E}_{q}A &=2J\sin(k) \big[-Ae^{-q}-Ae^{-q}\big]\nonumber \\
&= -4J\sin(k)e^{-q}\nonumber \\
&= 4J\sin(k)\sinh(q) - \mathcal{V}\nonumber \\
\end{align}
From this, we find that $q$ is
\begin{equation}
    q = \cosh^{-1}\bigg[ \frac{\mathcal{V}}{4J\sin(k)} \bigg]
\end{equation}
If now assume to be near the bottom of the band such that $\sin(k)\approx k$ then we find the energy and diffusion constant as follows,
\begin{align}
\label{eq:AppPotBSE}
    \mathcal{E}_{q} &= 4J\sin(k) \bigg[ \frac{\sqrt{\mathcal{V}^{2}-16J^{2}\sin^{2}(k)}}{4J\sin(k)}\bigg]-\mathcal{V} \approx \frac{-8J^{2}k^{2}}{\mathcal{V}}\nonumber \\
    \mathcal{D} &= \frac{8J^{2}}{\mathcal{V}}
\end{align}
Where because $k$ is small we binomially expand the square root and find that the bound state energy scales as $k^{2}$ which indicates diffusion at late times.

As $k$ increases, $q$ decreases, becoming zero at a critical $k$ given by $4J\sin k =\mathcal{V}$ for $\mathcal{V}<4J$, at which the energy equals -$\mathcal{V}$ and the bound state enters the continuum of scattering states and vanishes.
This transition manifests in the long-time behavior of the time-dependent correlation function by displaying increased oscillations before crossing over to diffusion; see Fig.~\ref{fig:BoundStateEnScalePot}(a) for numerical simulations of Eq.~\eqref{eq:AppSHNP} for different $\mathcal{V}/J$.
In Fig.~\ref{fig:BoundStateEnScalePot}(d), we compare our analytical result for $A_{\ell}$ to diagonalization of the eigenvalue equations, when $k=0.252$, $q\approx 0$ and increasing $k$ anymore cause the solution to crossover to a scattering state.
We also emphasize that the diffusion constant vanishes as $\mathcal{V}\rightarrow \infty$ indicating emergent localization.

\section{Diffusion Constant for Static Hopping with Bond and Onsite Noise.}

In this section, we give an explicit solution to the bound state energy in the case of finite noise with no static potential, i.e., $\gamma=0$ in Eq.~\eqref{eq:CoeffDiffEq}.
We begin with the eigenvalue equation,
\begin{equation}
\label{eq:AppSHBNPS}
    \mathcal{E}_{q}A_{\ell} = t_{k}\bigg[A_{\ell+1} - A_{\ell-1} \bigg] + \Gamma\bigg[-2A_{\ell}+\delta_{\ell,1}A_{-1}+\delta_{\ell,-1}A_{1}+2\cos(2k)\delta_{\ell,0}A_{0}\bigg] + \mathcal{V}\bigg[\delta_{\ell,0}-1\bigg]A_{\ell}.
\end{equation}
We have dropped the subscript $k$ since we are solving these for a fixed momentum and $t_{k}=2J\sin(k)$ is the effective hopping strength.
For finite noise, the above equation is non-hermitian and effectively describes the dynamics of a one-dimensional hopping model.
The above equation splits into four distinct eigenvalue equations,
\begin{align}
\label{eq:AppSHBNPSExpanded}
        \mathcal{E}_{q}A_{0} &= t_{k}\left(A_{1}-A_{-1} \right) -4\Gamma \sin^{2}(k)A_{0},\quad \mathcal{E}_{q}A_{1} = t_{k}\left(A_{2}-A_{0} \right) -\big[2\Gamma+\mathcal{V}\big] A_{1} + \Gamma A_{-1}\nonumber \\
        \mathcal{E}_{q}A_{-1} &= t_{k}\left(A_{0}-A_{-2} \right) -\big[2\Gamma+\mathcal{V}\big] A_{-1} + \Gamma A_{1},\quad \mathcal{E}_{q}A_{\ell} = t_{k}\left(A_{\ell+1}-A_{\ell-1} \right) -\big[2\Gamma+\mathcal{V}\big] A_{\ell}.
\end{align}
The final equation describes the bulk dynamics and sets the energy of the problem, while the first three equations are boundary conditions that constrain the allowed values that $q$ can take.
The three remaining equations solve for the variables $A_{1}/A_{-1}$, $A_{0}/A_{-1}$, and $q$.
Using the same ansatz as the previous two sections, leads to the energy,
\begin{equation}
    \label{eq:AppSHNHNP}
    \mathcal{E}_{q} = 4J\sin(k)\sinh(q) - \big[\mathcal{V}+2\Gamma \big]
\end{equation}
The middle two equations solve for the ratios, and the equation for $A_{0}$ restricts the values $q$ takes.
First, rewrite the middle two equations,
\begin{align}
    \bigg[ \mathcal{E}_{q} + 2Je^{-q}\sin(k) + 2\Gamma +\mathcal{V} \bigg]\frac{A_{1}}{A_{-1}} + 2J\sin(k)\frac{A_{0}}{A_{-1}} = \Gamma\\
    \bigg[ \mathcal{E}_{q} + 2Je^{-q}\sin(k) + 2\Gamma +\mathcal{V} \bigg] - 2J\sin(k)\frac{A_{0}}{A_{-1}} = \Gamma\frac{A_{1}}{A_{-1}}
\end{align}
Solving these equations yields the same consistency equations as above,
\begin{equation}
\label{eq:RatioConstraint2}
    \frac{A_{1}}{A_{-1}} = -1,\quad \frac{A_{0}}{A_{-1}} = e^{q} + \frac{\Gamma }{2J\sin(k)}
\end{equation}
Here $A_{1}$ and $A_{-1}$ are determined through normalizing $A_{\ell}$. The value of $A_{0}$ is determined through the second equation.
The ratios above, combined with the final eigenvalue equation for $A_{0}$ leads to the constraint,
\begin{equation}
\label{eq:QconstraintEquation2}
    \bigg[\mathcal{E}_{q}+4\Gamma \sin^{2}(k) \bigg] \bigg[ e^{q} + \frac{\Gamma }{2J\sin(k)}\bigg] = -4J\sin(k)
\end{equation}
The above equation is a cubic equation for $q$, which can be solved exactly.
We find that the bound state energy has the scaling,
\begin{align}
\label{eq:FiniteNoiseEnergyApp2}
    \mathcal{E}_{q} &= -4\bigg[\Gamma + \frac{2J^{2}}{\mathcal{V}+3\Gamma}\bigg]k^{2}\nonumber \\
    \mathcal{D} &= 4\bigg[\Gamma + \frac{2J^{2}}{\mathcal{V}+3\Gamma}\bigg]
\end{align}
The above solution in the limit $\mathcal{V}=0$ becomes Eq.~\eqref{eq:FiniteNoiseEnergyApp} for the case of static hopping with bond noise, while in the limit $\Gamma=0$ the bound state energy becomes Eq.~\eqref{eq:AppPotBSE} for the case of static hopping with onsite noise.
Notice, that the diffusion constant will generically have non-monotonic behavior where when $\Gamma$ is increasing there will be a regime of suppressed diffusion from the second term, while when $\Gamma$ continues to increase the first term will dominate leading to monotonic growth to the asymptotic value $4\Gamma$.

\section{Diffusion Constant for Linear Potential with Bond Noise.}

\begin{figure}[t]
    \includegraphics[width=\textwidth]{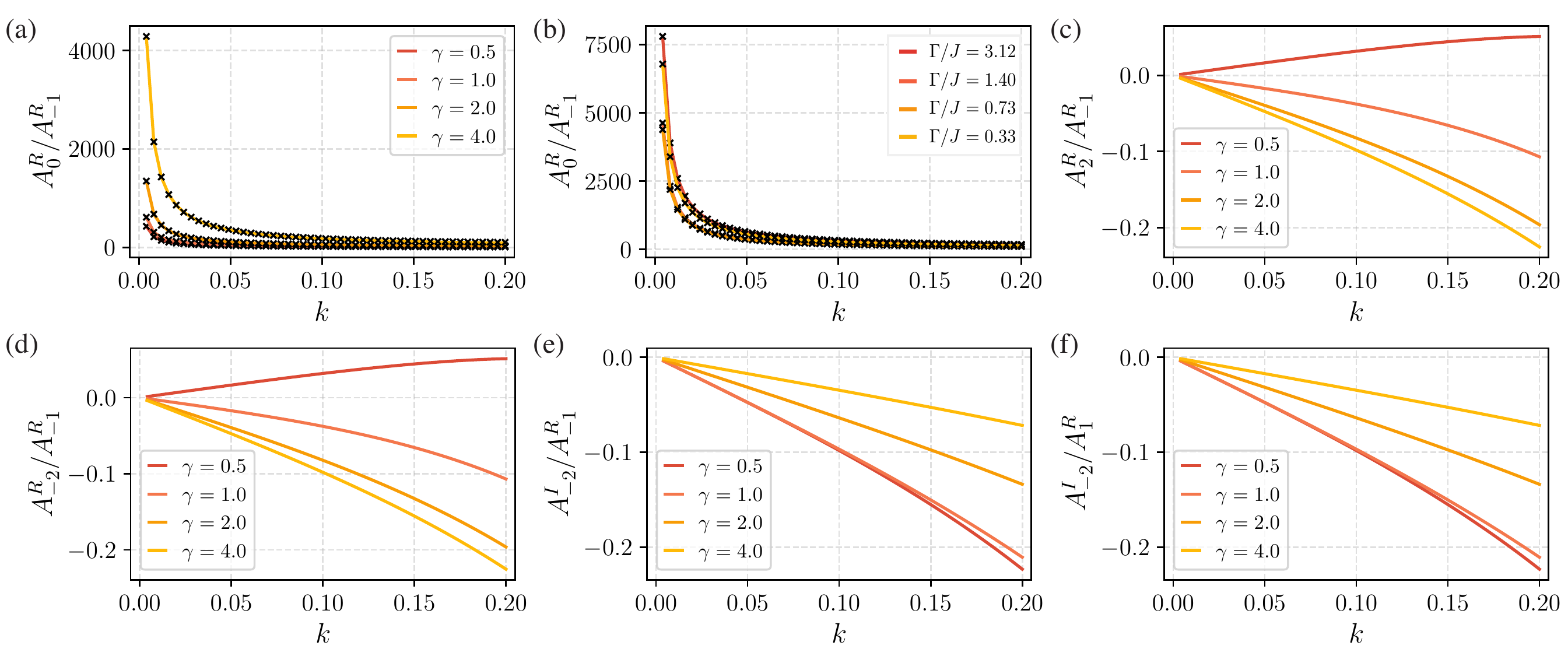}
    \caption{\textit{Coefficient Ratios for Linear Potential.}
    (a) Coefficient ratio $A^{R}_{0}/A^{R}_{-1}$, which governs the bound state energy scaling, as a function of momentum $k$ for different potential strengths and fixed $\Gamma/J=1$. The black crosses are from numerical diagonalization of Eq.~\eqref{eq:FiniteNoiseEigValStark}, while the curves are analytical results.
    (b) Same as (a) with fixed potential strength $\gamma=4$ and different noise strengths. It is important to note that in both (a) and (b) the final points at $k=0.2$ are $A^{R}_{0}/A^{R}_{-1}>100$.
    (c) $A^{R}_{2}/A^{R}_{-1}$, (d) $A^{R}_{-2}/A^{R}_{-1}$, (e) $A^{I}_{-2}/A^{R}_{-1}$, (f) $A^{I}_{-2}/A^{R}_{1}$ for different potential strengths.
    To determine the analytical form of the bound state energy Eq.~\eqref{eq:StarkBoundEnergy} we neglected all $A^{R(I)}_{\pm 2}$ terms in the operator equations by arguing they are small compared to $A^{R}_{0}/A^{R}_{-1}$. These plots provide a numerical justification of this approximation, where clearly the ratios (c)-(f) are significantly smaller over the range of momentum and give a negligent contribution to the energy.
    Parameters: $N=400$
    }
    \label{fig:RatioComparisons}
\end{figure}

We now consider the case where we include a static potential in the operator equation. Specifically, we consider a linear potential whereby the system undergoes Stark localization, and transport is absent in the clean limit.
The eigenvalue equation has the form,
\begin{equation}
\label{eq:FiniteNoiseEigValStark}
    \mathcal{E}_{q}A_{\ell} = 2J\sin(k) \bigg[A_{\ell+1} - A_{\ell-1} \bigg] + \Gamma\bigg[-2A_{\ell}+\delta_{\ell,1}A_{-1}+\delta_{\ell,-1}A_{1}+2\cos(2k)\delta_{\ell,0}A_{0}\bigg]+i\gamma \ell A_{\ell}.
\end{equation}
We note that with this potential the coordinate $\mathcal{R}$ still has spatially translation invariance, however the explicit dependence on $\ell$ prevents the ansatz used above.
Moreover, we split Eq.~\eqref{eq:FiniteNoiseEigValStark} into boundary and bulk equations,
\begin{align}
\label{eq:FiniteNoiseEquationsStark}
        \mathcal{E}_{q}A_{0} &= t_{k}\left(A_{1}-A_{-1} \right) -4\Gamma\sin^{2}(k)A_{0},\quad \mathcal{E}_{q}A_{1} = t_{k}\left(A_{2}-A_{0} \right) -2\Gamma A_{1} + \Gamma A_{-1}+i\gamma A_{1}\nonumber \\
        \mathcal{E}_{q}A_{-1} &= t_{k}\left(A_{0}-A_{-2} \right) -2\Gamma A_{-1} + \Gamma A_{1}-i\gamma A_{-1},\quad \mathcal{E}_{q}A_{\ell} = t_{k}\left(A_{\ell+1}-A_{\ell-1} \right)-2\Gamma A_{\ell}+i\gamma \ell A_{\ell}.
\end{align}
The bulk equation without the boundary conditions is anti-hermitian; therefore, the energies are imaginary with a fixed real part, i.e., $\mathcal{E}_{q}=-2\Gamma+i\gamma q$ where $q\in \{-\ell_{\text{max}}, \ell_{\text{max}}\}$.
The energy becomes purely imaginary in the noise-free limit, forming a Wannier-Stark ladder leading to non-thermal Bloch oscillations.
Including the boundary conditions $A_{0}, A_{1}, A_{-1}$ at finite noise has the effect of turning the set of equations non-hermitian, causing an eigenstate to emerge from the ladder with purely real energy, which is the bound state that governs the hydrodynamic behavior.
The bulk equation is the modified Bessel function recurrence relation, solved with the following,
\begin{equation}
    A_{\ell} = \begin{cases}
    A\mathcal{I}_{\nu_{-}}(-2it_{k}/\gamma)& \text{if}\ \ell \leq -1\\
    \\
    B\mathcal{I}_{\nu_{+}}(-2it_{k}/\gamma)& \text{if}\ \ell \geq 1.
\end{cases}
\end{equation}
where $\mathcal{I}_{\nu_{\pm}}(iz)$ is the modified Bessel function of the first kind of order $\nu_{\pm}=\frac{i(\mathcal{E}_{q}+2\Gamma)}{\gamma}\pm \ell$.
The difference in the sign for the Bessel function order ensures that the solution decays to zero as $\ell\rightarrow \pm \infty$; otherwise, it is generically a poorly behaved function on the left half.

\subsubsection{Approximate solution to Bound State Energy.}
Unlike the case without a potential, here, determining the analytical form of the bound state energy is tricky because the solution to the recurrence relation explicitly depends on the energy through the Bessel function order. In the previous section, determining the energy came down to finding a particular ratio of the coefficients.
In this section, we construct a solution to the bound state energy by making a few approximations that make the calculation more tenable.
To begin, we expand the three equations for $A_{0}$ and $A_{\pm 1}$ into six equations by breaking them into real and imaginary parts, i.e., $A_{\ell}=A^{\mathbb{R}}_{\ell}+iA^{I}_{\ell}$.
The following set is as follows,
\begin{align}
\label{eq:FiniteNoiseEquationsStarkExpand}
        \mathcal{E}_{q}A^{\mathbb{R}}_{0} &= t_{k}\left(A^{\mathbb{R}}_{1}-A^{\mathbb{R}}_{-1} \right)-4\Gamma \sin^{2}(k)A^{\mathbb{R}}_{0},\quad \mathcal{E}_{q}A^{\mathbb{R}}_{1} = t_{k}\left(A^{\mathbb{R}}_{2}-A^{\mathbb{R}}_{0} \right)-2\Gamma A^{\mathbb{R}}_{1} + \Gamma A^{\mathbb{R}}_{-1} - \gamma A^{I}_{1}\nonumber \\
        \mathcal{E}_{q}A^{\mathbb{R}}_{-1} &= t_{k}\left(A^{\mathbb{R}}_{0}-A^{\mathbb{R}}_{-2} \right)-2\Gamma A^{\mathbb{R}}_{-1} + \Gamma A^{\mathbb{R}}_{1}+\gamma A^{I}_{-1},\quad \mathcal{E}_{q}A^{I}_{0} = t_{k}\left(A^{I}_{1}-A^{I}_{-1} \right)-4\Gamma \sin^{2}(k)A^{I}_{0}\nonumber \\
        \mathcal{E}_{q}A^{I}_{1} &= t_{k}\left(A^{I}_{2}-A^{I}_{0} \right)-2\Gamma A^{I}_{1} + \Gamma A^{I}_{-1} +\gamma A^{\mathbb{R}}_{1},\quad \mathcal{E}_{q}A^{I}_{-1} = t_{k}\left(A^{I}_{0}-A^{I}_{-2} \right)-2\Gamma A^{I}_{-1} + \Gamma A^{I}_{1}-\gamma A^{\mathbb{R}}_{-1}.
\end{align}
From numerics we know the following features: $A^{\mathbb{R}}_{1}/A^{\mathbb{R}}_{-1}=-1$, $A^{I}_{1}/A^{I}_{-1}=1$, and $A^{I}_{0}=0$.
Moreover, the bound state energy is solely determined by the ratio $A^{\mathbb{R}}_{0}/A^{\mathbb{R}}_{-1}$,
\begin{equation}
\label{eq:StarkBoundEn}
    \bigg[\mathcal{E}_{q}+4\Gamma \sin^{2}(k)\bigg]\frac{A^{\mathbb{R}}_{0}}{A^{\mathbb{R}}_{-1}} = -2t_{k}
\end{equation}
To solve the above set of equations, we first study the ratio $A^{\mathbb{R}}_{0}/A^{\mathbb{R}}_{-1}$ as a function of momentum for different $\gamma$ and $\theta$ in Fig.~\ref{fig:RatioComparisons}(a) and (b) where the functional form roughly decays as $\sim 1/k$.
We then numerically compare the ratios including $A^{\mathbb{R}(I)}_{\pm 2}$ as a function of momentum in Fig.~\ref{fig:RatioComparisons}(c)-(f) and find that they are small compared to $A^{\mathbb{R}}_{0}/A^{\mathbb{R}}_{-1}$ which has a final value greater than one-hundred.
Moreover, in the equations above we set all $A^{\mathbb{R}(I)}_{\pm 2}=0$.
Next, we use the imaginary equations to determine the ratio $A^{I}_{1}/A^{\mathbb{R}}_{-1}$ and $A^{I}_{-1}/A^{\mathbb{R}}_{-1}$, which we find to be,
\begin{equation}
    \frac{A^{I}_{-1}}{A^{\mathbb{R}}_{-1}} = \frac{-\gamma}{\mathcal{E}_{q}+\Gamma}.
\end{equation}
Inserting this relation into the equations for $A^{\mathbb{R}}_{1}$,$A^{\mathbb{R}}_{-1}$ we find that,
\begin{equation}
    \frac{A^{\mathbb{R}}_{0}}{A^{\mathbb{R}}_{-1}} = \frac{\gamma^{2} +3\Gamma^{2}}{2J\Gamma \sin(k)}
\end{equation}
Inserting into Eq.~\eqref{eq:StarkBoundEn} will give a cubic equation in $\mathcal{E}_{q}$ just as in the previous section.
Fig.~\ref{fig:RatioComparisons}(a) and (b) we plot the above analytical result~[solid curves] neglecting terms with $\mathcal{E}_{q}$ and find perfect agreement with numerical diagonalization~[black crosses].
With this approximation, we find that the bound state energy has the scaling,
\begin{align}
\label{eq:StarkBoundEnergy}
    \mathcal{E}_{q}&=-4\Gamma \sin^{2}(k) \bigg[\frac{2J^{2} +\gamma^{2}+3\Gamma^{2}}{\gamma^{2}+3\Gamma^{2}} \bigg]\approx -4\Gamma \bigg[\frac{2J^{2} +\gamma^{2}+3\Gamma^{2}}{\gamma^{2}+3\Gamma^{2}} \bigg]k^{2}\nonumber \\
    \mathcal{D} &= 4\Gamma \bigg[\frac{2J^{2} +\gamma^{2}+3\Gamma^{2}}{\gamma^{2}+3\Gamma^{2}} \bigg]
\end{align}
Note, the above result preserves the feature that when $\Gamma=0$, $\mathcal{E}_{q}=0$ because there is not a bound state with purely real energy, but rather an entire spectrum of equally spaced bound states with purely complex energies from the emergent Stark ladder.
If instead we take the limit, $\gamma=0$; we recover the bound state energy for static hopping with bond noise.
We point out that in this case, when momentum increases, the diffusive bound state does not meet a continuum of scattering states but rather other bound states from the localization.
Interestingly, these bound states do not appear in the correlation function, indicating that the ground state we found is significantly more dominant.

\section{Diffusion Constant for Linear Potential with Onsite Noise.}
In this section, we study a similar model as above except with the noise on each site by coupling to the density rather than the hopping, i.e., $\Gamma = 0$.
Moreover, the boundary and bulk equations have the form,
\begin{align}
\label{eq:FiniteNoiseEquationsStark2}
        \mathcal{E}_{q}A_{0} &= t_{k}\left(A_{1}-A_{-1} \right),\quad \mathcal{E}_{q}A_{1} = t_{k}\left(A_{2}-A_{0} \right) +\big[ i\gamma-\mathcal{V}\big] A_{1}\nonumber \\
        \mathcal{E}_{q}A_{-1} &= t_{k}\left(A_{0}-A_{-2} \right) -\big[ i\gamma+\mathcal{V}\big]A_{-1},\quad \mathcal{E}_{q}A_{\ell} = t_{k}\left(A_{\ell+1}-A_{\ell-1} \right)+\big[ i\gamma\ell-\mathcal{V}\big]A_{\ell}.
\end{align}
To elucidate the bound state energy scaling, we follow the same method in the previous section where we first split the above equations into real and imaginary parts and use the relations, $A^{\mathbb{R}}_{1}/A^{\mathbb{R}}_{-1}=-1$, $A^{I}_{1}/A^{I}_{-1}=1$, and $A^{I}_{0}=0$.
Next, we use the imaginary equations to determine the ratio $A^{I}_{1}/A^{\mathbb{R}}_{-1}$ and $A^{I}_{-1}/A^{\mathbb{R}}_{-1}$, which we find to be,
\begin{equation}
    \frac{A^{I}_{-1}}{A^{\mathbb{R}}_{-1}} = \frac{-\gamma}{\mathcal{E}_{q}+\mathcal{V}}.
\end{equation}
Inserting this relation into the equations for $A^{\mathbb{R}}_{1}$,$A^{\mathbb{R}}_{-1}$ we find that,
\begin{equation}
    \frac{A^{\mathbb{R}}_{0}}{A^{\mathbb{R}}_{-1}} = \frac{\gamma^{2}+\mathcal{V}^{2}}{\mathcal{V}t_{k}}
\end{equation}
where we have ignored terms with $\mathcal{E}_{q}$.
With this approximation, we find the bound state energy has the scaling,
\begin{align}
\label{eq:StarkBoundEnergy2}
    \mathcal{E}_{q}&=\bigg[\frac{-8J^{2}\mathcal{V}}{\gamma^{2}+\mathcal{V}^{2}}\bigg]k^{2}\nonumber \\
    \mathcal{D} &= \bigg[\frac{8J^{2}\mathcal{V}}{\gamma^{2}+\mathcal{V}^{2}}\bigg]
\end{align}
Note, the above result preserves the feature that when $\mathcal{V}=0$, $\mathcal{E}_{q}=0$ because there is not a bound state with a purely real energy, but rather an entire spectrum of bound states with purely complex energies that are equally spaced from the emergent Stark ladder.
Moreover, if $\gamma = 0$ then we recover the result, $\mathcal{E}_{q}=\frac{-8J^{2}}{\mathcal{V}}k^{2}$ discussed in the main text of static hopping with onsite noise.

\section{Diffusion Constant for General Operator Equation of Motion.}

In this section, we present a derivation for the diffusion constant for Eq.~\eqref{eq:CoeffDiffEq} with all all terms contributing which is one of the primary results in the main-text.
Moreover, for this case the model effectively maps to a one-dimensional non-hermitian hopping model in an imaginary linear potential with delta function potential, the eigenvalue equation becomes,
\begin{align}
\label{eq:FullEigenvalueEqns}
        \mathcal{E}_{q}A_{0} &= t_{k}\left(A_{1}-A_{-1} \right) -4\Gamma\sin^{2}(k)A_{0},\quad \mathcal{E}_{q}A_{1} = t_{k}\left(A_{2}-A_{0} \right) -2\Gamma A_{1} + \Gamma A_{-1}+i\gamma A_{1}-\mathcal{V}A_{1}\nonumber \\
        \mathcal{E}_{q}A_{-1} &= t_{k}\left(A_{0}-A_{-2} \right) -2\Gamma A_{-1} + \Gamma A_{1}-i\gamma A_{-1}-\mathcal{V}A_{-1},\quad \mathcal{E}_{q}A_{\ell} = t_{k}\left(A_{\ell+1}-A_{\ell-1} \right)-2\Gamma)A_{\ell}+i\gamma \ell A_{\ell} -\mathcal{V}A_{\ell}.
\end{align}
Just like in the previous two sections, we split the above eigenvalue equations into their respective real and imaginary parts and utilize the relations $A^{\mathbb{R}}_{1}/A^{\mathbb{R}}_{-1}=-1$, $A^{I}_{1}/A^{I}_{-1}=1$, and $A^{I}_{0}=0$ to simplify the equations.
Using the imaginary equations for $A_{\pm}$, we determine the following ratio,
\begin{equation}
    \frac{A^{I}_{-1}}{A^{\mathbb{R}}_{-1}} = \frac{-\gamma}{\mathcal{E}_{q}+\mathcal{V}+\Gamma}.
\end{equation}
This result then leads to ratio,
\begin{equation}
    \frac{A^{\mathbb{R}}_{0}}{A^{\mathbb{R}}_{-1}} = \frac{\gamma^{2} +(3\Gamma + V)(\Gamma + V)}{2J\sin(k)(\Gamma + V)}
\end{equation}
where we have ignored terms with $\mathcal{E}_{q}$.
With this approximation, we find the bound state energy has the scaling,
\begin{align}
\label{eq:FullBoundEnergy}
    \mathcal{E}_{q}&=-8\bigg[\frac{\Gamma}{2}+\frac{J^{2}(\mathcal{V}+\Gamma)}{\gamma^{2}+(3\Gamma + V)(\Gamma + V)}\bigg]k^{2}\nonumber \\
    \mathcal{D} &= 8\bigg[\frac{\Gamma}{2}+\frac{J^{2}(\mathcal{V}+\Gamma)}{\gamma^{2}+(3\Gamma + V)(\Gamma + V)}\bigg]
\end{align}

\section{Numerical Methods for Determining the Coefficients.}
We now outline how in the main text we numerically determined the two-point correlation function from the coefficient, $A_{\ell, k}$.
First, we put an upper limit on the operator length labelled $\ell_{\text{max}}$ such that,
\begin{equation}
\Vec{A} = \begin{pmatrix}A_{-\ell_{\text{max}},k},\ldots, & A_{0,k},\ldots, & A_{\ell_{\text{max}}, k}\end{pmatrix}.
\end{equation}
The above vector obeys the matrix differential equation,
\begin{equation}
    \frac{d\Vec{A}(t)}{dt}=M\Vec{A}(t)
\end{equation}
where the matrix $M$ is determined from Eq.~(10) in the main text which is diagonalized as $M=VDV^{-1}$ where $V$ is set of eigenvectors and $D=\text{diag}(\lambda_{1},\ldots,\lambda_{2\ell_{\text{max}}+1})$.
Upon making the above change of basis the differential equation above becomes,
\begin{equation}
    \frac{d\Vec{\Sigma}(t)}{dt}=D\Vec{\Sigma}(t).
\end{equation}
Here $\Vec{\Sigma}=V^{-1}\Vec{A}$ and the equation is easily integrated to give, $\Vec{\Sigma}(t)=e^{Dt}\Vec{\Sigma}(0)$.
Mapping back to the original basis, the coefficients are found to be,
\begin{equation}
    \Vec{A}(t)=\left(Ve^{Dt}V^{-1}\right)\Vec{A}(0).
\end{equation}
We then extract the element $[\Vec{A}(t)]_{0,k}$ for each momentum and then perform the following sum,
\begin{equation}
    C_{x,y}(t)=\frac{1}{8\pi}\sum_{k}[\Vec{A}(t)]_{0,k} e^{i(x-y)k}.
\end{equation}

\bibliography{references}